\documentclass[%
reprint,
 amsmath,amssymb,
 aps,
prb,
floatfix
]{revtex4-2}

\usepackage{graphicx}
\usepackage{dcolumn}
\usepackage{bm}
\usepackage{hyperref}

\usepackage{enumitem}

\begin{document}


\title{Bridging the continuum and the kinetic-Boltzmann theories of heat flow through generalized Knudsen numbers}

\author{Nikhil Malviya}
\author{Navaneetha K. Ravichandran}
\email{navaneeth@iisc.ac.in}
\affiliation{%
 Department of Mechanical Engineering, Indian Institute of Science, Bangalore 560012, India
 }%

\date{\today}

\begin{abstract}
Heat conduction in non-magnetic semiconductor crystals is fundamentally governed by the linearized Peierls-Boltzmann equation (LPBE) for phonon transport, that arises out of a kinetic theory for phonon quasiparticles. Yet, continuum theories such as the Fourier's heat diffusion law and the non-Fourier hydrodynamic heat equation are often used to explain the experimental observations of heat flow in these material systems. Here, we show that formal reductions of the phonon quasiparticle transport theory into such equivalent continuum descriptions are possible only for the limiting values of a set of generalized Knudsen numbers defined on the eigenspectra of the phonon collision operator ($\bm{\Omega}$). We arrive at these generalized Knudsen numbers by systematically reducing the governing equation for phonon thermal transport in semiconductors --- the linearized Peierls-Boltzmann equation (LPBE) --- in the eigenbasis of $\bm{\Omega}$, into the continuum equations for temperature dynamics corresponding to the Fourier-diffusive as well as the weakly quasiballistic and the hydrodynamic heat flow regimes. We further show that all of these heat flow regimes along with the ballistic heat flow regime can be described by a single continuum equation for the temperature field that originates from the eigenmode analysis of the LPBE, thus offering a unified picture of all possible heat flow regimes in non-magnetic semiconducting crystals. Using quantitative examples on twenty three technologically important semiconductors, we show that several previously-unidentified features of the non-Fourier heat flow regimes emerge from this framework of generalized Knudsen numbers such as (1) the mutual exclusivity of the weakly quasiballistic and the hydrodynamic heat flow regimes in any material, (2) a characteristic heating length for the strongest hydrodynamic heat flow signatures, (3) length-dependent velocity of the hydrodynamic second sound temperature wave, (4) characteristic frequency-domain temperature response distinguishing the hydrodynamic second sound and the ballistic heat flow regimes and, (5) a new non-oscillatory signature of transient hydrodynamic heat flow that has not been reported in the literature till now. Our work formally bridges the continuum descriptions of the Fourier-diffusive as well as the non-Fourier heat flow regimes with the microscopic particulate description of heat flow offered by the kinetic-Boltzmann theory for phonon transport, and provides insights into the important signatures of temperature dynamics in each of these heat flow regimes, that will aid in their unambiguous experimental observations in the future.
\end{abstract}

\maketitle

\clearpage

\section{Introduction} \label{sec:introduction}
The classical description of heat conduction in solids is based on the Fourier’s diffusion law, which relates the generated heat flux to the applied temperature gradient as a linear response, with the proportionality constant - the thermal conductivity ($\kappa$), described as a purely material property, independent of the geometry of the sample. The Fourier's law, in conjunction with the equation for energy conservation leads to a parabolic diffusion equation for the continuum temperature field --- the Fourier heat equation (FHE), expressed for homogeneous materials as:
\begin{equation}
    \frac{\partial \Delta T}{\partial t} - \rho \frac{\partial^{2} \Delta T}{\partial x^{2}} = 0
    \label{eq:fourier_diffusion_heat_equation}
\end{equation}
where $\Delta T \equiv \Delta T \left( x, t \right) = T \left( x, t \right) - T_{0}$ is the deviation of the spatio-temporal temperature profile from the global equilibrium temperature ($T_{0}$) at time ($t$) and position ($x$), and $\rho = \kappa / C_{0}$ is the thermal diffusivity of the material with the heat capacity $C_{0}$. Recent heat flow experiments have demonstrated a deviation from the Fourier's law with $\kappa$ becoming a length-dependent quantity at micro- to nano-scales, transitioning from diffusive to weakly quasiballistic~\cite{johnson_direct_2013, ravichandran_spectrally_2018} or ballistic regimes~\cite{maire_ballistic_2017}. Additionally, in ultrahigh-$\kappa$ materials, heat flow can enter into a second sound regime with wave-like spatio-temporal oscillations of the temperature field~\cite{mcnelly_heat_1970, jackson_second_1970, jackson_thermal_1971, huberman_observation_2019, ding_observation_2022, huang_graphite_2024, xie_room-temperature_2026} -- a feature that is atypical of a governing parabolic differential equation [Eq.~\ref{eq:fourier_diffusion_heat_equation}].

These non-Fourier heat flow regimes are often described by other continuum equations on the temperature field similar to the FHE. For example, the weakly quasiballistic heat flow is described using a parabolic equation similar to the FHE called the weakly quasiballistic heat equation (wQHE) with $\rho$ replaced by a suppressed, length-dependent effective diffusivity ($\rho_{s}$)~\cite{minnich_quasiballistic_2011}, given by:
\begin{align}
     \frac{\partial \Delta T}{\partial t} - \rho_{s} \frac{\partial^{2} \Delta T}{\partial x^{2}} = 0
    \label{eq:quasiballistic_heat_equation}
\end{align}
Similarly, the second sound heat flow regime follows a hyperbolic, damped wave equation called the hydrodynamic heat equation (HHE)~\cite{hardy_phonon_1970}, given by:
\begin{align}
        \frac{\partial^{2} \Delta T}{\partial t^{2}} + \mu \frac{\partial \Delta T}{\partial t} - v_{ud}^{2} \frac{\partial^{2} \Delta T}{\partial x^{2}} = 0
    \label{eq:hyperbolic_heat_equation}
\end{align}
where $\mu$ is the damping coefficient and $v_{ud}$ is the undamped velocity of the temperature oscillations.

These continuum descriptions are valuable because the temperature field equations are straightforward to solve, thus allowing analysis of non-Fourier heat flow regimes even in intricate nanoscale geometries~\cite{johnson_direct_2013, ravichandran_spectrally_2018, huberman_observation_2019, ding_observation_2022, beardo_observation_2021}. However, it is not known apriori if the heat flow regime in a semiconductor device will be Fourier-diffusive, weakly quasiballistic or hydrodynamic under a specified experimental condition. In other words, there is no rigorous established way of anticipating when the FHE, the wQHE or the HHE must be used to describe the heat flow. These predictions cannot be gleaned convincingly from available experimental measurements as well. For example, while only weakly quasiballistic heat flow has been observed in silicon around and below room temperature~\cite{ravichandran_spectrally_2018, johnson_direct_2013, minnich_thermal_2011}, the hydrodynamic heat flow regime has been observed only in graphite~\cite{huberman_observation_2019, ding_observation_2022, xie_room-temperature_2026} and germanium~\cite{beardo_observation_2021} beyond 100~K. Even within graphite, the hydrodynamic heat flow regime is not observed in thinner samples~\cite{jeon_grain_2025}.

In this work, we address this challenge by deriving the necessary conditions for Fourier-diffusive, weakly quasiballistic and hydrodynamic heat flow regimes [Eqs. ~\ref{eq:fourier_diffusion_heat_equation},~\ref{eq:quasiballistic_heat_equation} and~\ref{eq:hyperbolic_heat_equation}] from the properties of the solutions of the governing equation for heat flow in semiconductors - the linearized Peierls-Boltzmann equation (LPBE) for phonon transport, given by: 
\begin{align}
    \frac{\partial f'_{\lambda}}{\partial t} + \mathbf{v}_{\lambda} \cdot \bm{\nabla} f'_{\lambda} = & - \sum_{\lambda_{1}} \Omega_{\lambda \lambda_{1}} f'_{\lambda_{1}} + \dot{\mathcal{H}}_{\lambda}
    \label{eq:phonon_lpbe_time_spatial_domain}
\end{align}
Here, $f'_{\lambda} $ is the linearized, deviational non-equilibrium phonon distribution function at time ($t$) and position ($\mathbf{x}$), with wave vector $\mathbf{q}$ and polarization $j$ (labeled jointly as $\lambda = \{ \mathbf{q}, j\}$), $\mathbf{v}_{\lambda}$ is the phonon group velocity, $\bm{\Omega}$ is the collision matrix, and $\dot{\mathcal{H}}_{\lambda}$ is the phonon-specific heat source. The LPBE is the fundamental governing equation for phonon transport whenever the phonon quasiparticle picture holds~\cite{ravichandran_unified_2018} and is general enough to describe the Fourier-diffusive, the weakly quasiballistic and the hydrodynamic heat flow regimes unlike the regime-specific temperature field equations [Eqs.~\ref{eq:fourier_diffusion_heat_equation},~\ref{eq:quasiballistic_heat_equation} and~\ref{eq:hyperbolic_heat_equation}] introduced earlier, as well as the ballistic heat flow regime that could occur when phonons transport heat without undergoing collisions. 

The linearity of Eq.~\ref{eq:phonon_lpbe_time_spatial_domain} and the symmetric nature of $\bm{\Omega}$ allow for a solution of Eq.~\ref{eq:phonon_lpbe_time_spatial_domain} in terms of the complete orthonormal eigenbasis of $\bm{\Omega}$. Taking advantage of this feature, we draw inspiration from Ref.~\cite{hardy_phonon_1970} to derive the conditions that activate the Fourier-diffusive and different non-Fourier regimes in terms of generalized \textit{Knudsen numbers} that depend on the properties of the eigenmodes of $\bm{\Omega}$. We show that these generalized Knudsen numbers unveil several important characteristics of these unconventional heat flow regimes that have not been previously reported --- specifically (1) the mutual exclusivity of weakly quasiballistic and hydrodynamic heat flow regimes in a material, (2) a length-dependent velocity of hydrodynamic second sound temperature wave and a characteristic heating length for the strongest hydrodynamic second sound oscillations, (3) experimental signatures of the spatio-temporal temperature field variations that conclusively distinguish between the hydrodynamic second sound and the ballistic heat flow regimes, (4) a previously-unidentified transient hydrodynamic regime without second sound-like oscillations and, (5) a generalized suppression function that quantifies the effective $\kappa$ as a function of heating length in the weakly quasiballistic regime, that is applicable even when the relaxation time approximation (RTA) of the LPBE fails. Finally, we demonstrate the predictive power of the conditions on these generalized Knudsen numbers by identifying, from first principles, experimental conditions under which such Fourier-diffusive and non-Fourier heat flow regimes can be realized in twenty two different cubic semiconductors as well as in graphene, where we find the strongest hydrodynamic signatures even at temperatures as high as 100~K. Our work lays down a rigorous foundational connect between the continuum and the kinetic theories of heat flow, that will aid in predicting and experimentally realizing unconventional heat flow regimes in semiconductors that transcend the textbook Fourier's law of heat diffusion.

\section{Generalized heat equation from the LPBE} \label{sec:generalized_heat_equations}
To derive the generalized Knudsen numbers as well as the necessary conditions on them for observing Fourier-diffusive and non-Fourier heat flow regimes, we begin by systematically reducing the LPBE into the respective continuum temperature field equations. The LPBE [Eq.~\ref{eq:phonon_lpbe_time_spatial_domain}] is a coupled set of linear first-order partial differential equations, with the coupling originating from the collision term $\sum_{\lambda_1}\Omega_{\lambda\lambda_1}f'_{\lambda_1}$. Following Ref.~\cite{hardy_phonon_1970}, first, we express the vectors $f'_\lambda$ and $\dot{\mathcal{H}}_\lambda$ as linear combinations of the complete orthonormal eigenvectors of the symmetric matrix $\bm{\Omega}$. In this basis, Eq.~\ref{eq:phonon_lpbe_time_spatial_domain} becomes~\cite{hardy_phonon_1970, cepellotti_thermal_2016, hua_space-time_2020, malviya_efficient_2025}:
\begin{equation}
    \frac{\partial \vartheta^{\beta}}{\partial t} + \sum_{\alpha} \bm{\mathcal{V}}^{\beta \alpha} \cdot \bm{\nabla} \vartheta^{\alpha} = - \vartheta^{\beta} \sigma^{\beta} + h^{\beta}
    \label{eq:relaxon_lpbe_time_spatial_domain}
\end{equation}
where $f'_{\lambda} = \sum_{\beta} \vartheta^{\beta}  \mathfrak{e}_{\lambda}^{\beta}$ and $\dot{\mathcal{H}}_{\lambda} = \sum_{\beta} h^{\beta} \mathfrak{e}_{\lambda}^{\beta}$, with $\mathfrak{e}^{\beta}$ and $\sigma^{\beta}$ being the $\beta^{\text{th}}$ eigenvector and eigenvalue of $\bm{\Omega}$ respectively, with $\sigma^\beta \ge 0$ since $\bm{\Omega}$ is positive semi-definite, and $\bm{\mathcal{V}}^{\beta \alpha} = \sum_{\lambda} \mathfrak{e}_{\lambda}^{\beta} \mathbf{v}_{\lambda} \mathfrak{e}_{\lambda}^{\alpha}$. These eigenmodes are also referred to as \emph{relaxons} in the literature~\cite{cepellotti_thermal_2016}. The collision matrix $\bm{\Omega}$ has a null vector $\mathfrak{e}_{\lambda}^{0} = \sqrt{f_{\lambda}^{0} \left( f_{\lambda}^{0} + 1 \right)} \left( \hbar \omega_{\lambda} \right)/\sqrt{V k_{B} T_{0}^{2} C_{0}}$ corresponding to a thermal equilibrium distribution, with $f_{\lambda}^{0}$ being the equilibrium distribution at a temperature $T_{0}$.

Applying spatial and temporal Fourier transforms to Eq.~\ref{eq:relaxon_lpbe_time_spatial_domain}, and assuming that the heat source adds thermal energy to each phonon mode at a rate proportional to its heat capacity, i.e., $\mathcal{\dot{H}}_\lambda = h^0\mathfrak{e}^0_\lambda$~\cite{hua_space-time_2020}, we get:
\begin{subequations}
\begin{gather}
    \beta\ne0:-i \eta \tilde{\vartheta}^{\beta} - i \sum_{\alpha > 0} \bm{\xi} \cdot \bm{\mathcal{V}}^{\beta \alpha} \tilde{\vartheta}^{\alpha} + \tilde{\vartheta}^{\beta} \sigma^{\beta} = i \bm{\xi} \cdot \bm{\mathcal{V}}^{\beta 0} \tilde{\vartheta}^{0}
    \label{eq:relaxon_lpbe_frequency_domain} \\
    -i \eta \tilde{\vartheta}^{0} - i \sum_{\alpha > 0} \bm{\xi} \cdot \bm{\mathcal{V}}^{0 \alpha} \tilde{\vartheta}^{\alpha} = \tilde{h}^{0}
    \label{eq:relaxon_lpbe_frequency_domain_0}
\end{gather}
\end{subequations}
where $\xi$ and $\eta$ are spatial and temporal Fourier variables respectively, and the tilde over the variables represents their Fourier transforms. For $\beta = 0$, Eq.~\ref{eq:relaxon_lpbe_time_spatial_domain} (and its Fourier transform [Eq.~\ref{eq:relaxon_lpbe_frequency_domain_0}]) represents the energy conservation equation~\cite{hua_space-time_2020, malviya_efficient_2025} and the corresponding coefficient $\vartheta^{0}$ can be related to the temperature deviation ($\Delta T$) by noting that the change in the energy of the system due to this deviation in the temperature is $\Delta E = C_{0} \Delta T = \sqrt{k_{B} T_{0}^{2} C_{0}/V} \vartheta^{0}$ with $V$ being the crystal volume. 

Further, the eigenvectors of $\bm{\Omega}$ (i.e., $\mathfrak{e}_{\lambda}^{\beta}$) can be chosen with even ($\mathfrak{e}_{\lambda}^{\beta} = \mathfrak{e}_{-\lambda}^{\beta}$) or odd ($\mathfrak{e}_{\lambda}^{\beta} = -\mathfrak{e}_{-\lambda}^{\beta}$) parity with respect to the phonon indices due to the even parity of $\bm{\Omega}$, i.e., $\Omega_{\lambda\lambda_{1}} = \Omega_{\left(-\lambda\right)\left(-\lambda_{1}\right)}$, where, $-\lambda \equiv \left[-\mathbf{q}, j\right]$~\cite{hardy_phonon_1970, hua_space-time_2020, malviya_efficient_2025}. Since $\mathfrak{e}_{\lambda}^{0}$ is an even eigenvector, and the phonon group velocity has odd parity with respect to the phonon index i.e. $\mathbf{v}_{\lambda} = - \mathbf{v}_{-\lambda}$, the velocity $\bm{\mathcal{V}}^{0 \beta} ( = \bm{\mathcal{V}}^{\beta 0} )$ of the $\beta^{\text{th}}$ eigenmode is non-zero for the odd eigenmodes only, and $\bm{\mathcal{V}}^{\beta \alpha}$ is non-zero only when the eigenmodes $\mathfrak{e}_{\lambda}^\alpha$ and $\mathfrak{e}_{\lambda}^\beta$ are of different parities.

With this classification of the eigenmodes of $\bm{\Omega}$, Eqs.~\ref{eq:relaxon_lpbe_frequency_domain} and~\ref{eq:relaxon_lpbe_frequency_domain_0} can be split for even and odd eigenmodes as~\cite{hardy_phonon_1970}:
\begin{subequations}
\begin{align}
    & -i \eta \tilde{\vartheta}^{\bar{\beta}} - i \sum_{\alpha > 0} \bm{\xi} \cdot \bm{\mathcal{V}}^{\bar{\beta} \alpha} \tilde{\vartheta}^{\alpha} + \tilde{\vartheta}^{\bar{\beta}} \sigma^{\bar{\beta}} = 0
    \label{eq:even_relaxon_lpbe_frequency_domain} \\
    & -i \eta \tilde{\vartheta}^{\beta} - i \sum_{\bar{\alpha} > 0} \bm{\xi} \cdot \bm{\mathcal{V}}^{\beta \bar{\alpha}} \tilde{\vartheta}^{\bar{\alpha}} + \tilde{\vartheta}^{\beta} \sigma^{\beta} = i \bm{\xi} \cdot \bm{\mathcal{V}}^{\beta 0} \left(\zeta \Delta \tilde{T}\right)
    \label{eq:odd_relaxon_lpbe_frequency_domain} \\
    & -i \eta \left(\zeta \Delta \tilde{T}\right) - i \sum_{\alpha > 0} \bm{\xi} \cdot \bm{\mathcal{V}}^{0 \alpha} \tilde{\vartheta}^{\alpha} = \tilde{h}^{0}
    \label{eq:even_odd_relaxon_lpbe_frequency_domain_0}
\end{align}
\end{subequations}
where the even and the odd eigenmodes are distinguished by indexing them with or without bars (i.e., $\bar{\beta}$ and $\beta$) respectively, and we replace $\tilde{\vartheta^{0}}$ with $\zeta \Delta \tilde{T}$, where $\zeta=\sqrt{C_{0}V/k_{B}T_{0}^{2}}$. For simplicity, we assume that the driving heat source, and therefore, the phonon transport is one-dimensional (say along the x-direction), i.e., $\bm{\xi} \sim \xi_x\hat{\bm{x}}$, moving forward. 

Eliminating $\tilde{\vartheta}^{\bar{\beta}}$ using Eq.~\ref{eq:even_relaxon_lpbe_frequency_domain}, Eq.~\ref{eq:odd_relaxon_lpbe_frequency_domain} becomes:
\begin{align*}
    \left( -i \eta + \sigma^{\beta} \right) \tilde{\vartheta}^{\beta} & = i \xi_x \mathcal{V}^{\beta 0}_x \zeta \Delta \tilde{T} - \sum_{\bar{\alpha}, \gamma > 0} \frac{ \left( \xi_x\mathcal{V}^{\beta \bar{\alpha}} _x\right) \left( \xi_x\mathcal{V}^{\bar{\alpha} \gamma}_x \right) }{\sigma^{\bar{\alpha}}-i \eta} \tilde{\vartheta}^{\gamma}
\end{align*}
which can be recast in a compact form as: $\sum_{\gamma>0} \Gamma^{\beta \gamma} \tilde{\vartheta}^{\gamma} = i \xi_{x} \mathcal{V}_{x}^{\beta 0} \zeta \Delta \tilde{T}$ with the matrix $\bm{\Gamma}$ given by:
\begin{align}
    \Gamma^{\beta \gamma} = \left[ \left( -i \eta + \sigma^{\beta} \right) \Delta_{\beta \gamma} + \sum_{\bar{\alpha}>0} \frac{ \left( \xi_x \mathcal{V}^{\beta \bar{\alpha}}_x \right) \left( \xi_x \mathcal{V}^{\bar{\alpha} \gamma}_x \right) }{\sigma^{\bar{\alpha}}-i \eta} \right]    \label{eq:Gamma_matrix}
\end{align}

The unknown odd coefficients, $\tilde{\vartheta}^{\alpha}$, can now be obtained as:
\begin{align}
\tilde{\vartheta}^{\alpha} = i \sum_{\beta > 0} \left[ \Gamma^{-1} \right]^{\alpha \beta} \xi_x \mathcal{V}^{\beta 0}_x \zeta \Delta \tilde{T}      \label{eq:theta_alpha_Gen_sol}
\end{align}
which upon substitution into Eq.~\ref{eq:even_odd_relaxon_lpbe_frequency_domain_0} results in the generalized heat equation (GHE) in the frequency domain:
\begin{align}
    -i \eta \Delta \tilde{T} - i \left[ \sum_{\alpha, \beta > 0} \xi_x \mathcal{V}^{0 \alpha}_x \left[ \Gamma^{-1} \right]^{\alpha \beta} \xi_x \mathcal{V}^{\beta 0}_x \right] \Delta \tilde{T} & = \frac{\tilde{h}^{0}}{\zeta}.
    \label{eq:lpbe_frequency_domain_temperature_response}
\end{align}

The spatio-temporal temperature response $\Delta T\left(\mathbf{x}, t\right)$ can be obtained by solving Eq.~\ref{eq:lpbe_frequency_domain_temperature_response} for $\Delta\tilde{T}\left(\xi, \eta\right)$ and performing an inverse Fourier transform in space and time. We emphasize that the GHE [Eq.~\ref{eq:lpbe_frequency_domain_temperature_response}] is a \emph{single} continuum field equation for the temperature response that has been derived from the LPBE without introducing any additional approximations, and so, can describe any heat flow regime that the LPBE admits. In particular, Eq.~\ref{eq:lpbe_frequency_domain_temperature_response} serves as a single master equation for the temperature field that describes the Fourier-diffusive as well as the weakly quasiballistic, the hydrodynamic and the ballistic heat flow regimes.

\section{Generalized Knudsen numbers for classifying heat flow regimes} \label{sec:generalized_regime_classifiers}
To identify different heat flow regimes admitted by the LPBE, the coefficients of $\Delta\tilde{T}$ in Eq.~\ref{eq:lpbe_frequency_domain_temperature_response} must allow it to be reduced to one of the continuum field equations introduced earlier [Eqs.~\ref{eq:fourier_diffusion_heat_equation}-~\ref{eq:hyperbolic_heat_equation}]. The challenge in reducing Eq.~\ref{eq:lpbe_frequency_domain_temperature_response} to one of the continuum equations lies in the inversion of the large, dense matrix $\mathbf{\Gamma}$; specifically, the absence of an analytical inverse of $\bm{\Gamma}$ makes any further simplification challenging. However, trivial simplifications are possible if $\bm{\Gamma}$ were a diagonal matrix. For example, it can be readily seen that, when $\Gamma^{\beta \gamma} = \sigma^{\beta} \Delta_{\beta \gamma}$, Eq.~\ref{eq:lpbe_frequency_domain_temperature_response} reduces to the FHE [Eq.~\ref{eq:fourier_diffusion_heat_equation}] in frequency domain. We can generalize this simplification by identifying conditions under which the off-diagonal part of $\mathbf{\Gamma}$ ($\mathbf{\Gamma}_{od}$) can be neglected compared to its diagonal part ($\mathbf{\Gamma}_{d}$) to get $\mathbf{\Gamma} \approx \mathbf{\Gamma}_{d}$. To achieve this simplification, we introduce a diagonal normalization to $\bm{\Gamma}$ as: $\sqrt{\bm{\Gamma}_d^{-1}}\bm{\Gamma}\sqrt{\bm{\Gamma}_d^{-1}} = \bm{I} + \sqrt{\bm{\Gamma}_d^{-1}}\bm{\Gamma}_{od}\sqrt{\bm{\Gamma}_d^{-1}}$, and note that $\bm{\Gamma}_{od}$ can be neglected when $\max_{\beta,\gamma\ne\beta}\left|\Gamma_{od}^{\beta \gamma}\right| \sqrt{\left|\left[ \Gamma_{d}^{-1} \right]^{\beta \beta}\right|\left|\left[ \Gamma_{d}^{-1} \right]^{\gamma \gamma}\right|} \ll 1$ for all $\beta, \gamma\ne\beta > 0$. As detailed in Appendix~\ref{app_subsec:diagonal_dominance_condition}, for this diagonal dominance condition (DDC) on $\bm{\Gamma}$, it is sufficient to satisfy the following requirement: 
\begin{align}
      \chi_{DDC} \left( \xi_{x} \right) = \max_{\beta, \gamma\ne\beta > 0} \sum_{\bar{\alpha}>0} \frac{\left| \left( \xi_{x} \mathcal{V}^{\beta \bar{\alpha}}_{x} \right) \left( \xi_{x} \mathcal{V}^{\bar{\alpha} \left( \gamma \neq \beta \right)}_{x} \right) \right|}{\displaystyle \sigma^{\bar{\alpha}} \sqrt{\sigma^{\beta}\sigma^{\gamma}} } \ll 1  \label{eq:diagonal_dominance_condition}
\end{align}
From $\chi_{DDC}$, a generalized \emph{spatial} Knudsen number for the eigenmodes of $\bm{\Omega}$ arises as:
\begin{align}
    \mathcal{K}_{\beta\gamma} = \xi_x\sqrt{\sum_{\bar{\alpha}>0}\left(\frac{\mathcal{V}^{\beta\bar{\alpha}}_{x}}{\sqrt{\sigma^{\bar{\alpha}}\sigma^\beta}}\right)\left(\frac{\mathcal{V}^{\gamma\bar{\alpha}}_{x}}{\sqrt{\sigma^{\bar{\alpha}}\sigma^\gamma}}\right)}
    \label{eq:knudsen_number_tensor}
\end{align}
and the condition on $\chi_{DDC}$ [Eq.~\ref{eq:diagonal_dominance_condition}] becomes:
\begin{align}
    \chi_{DDC} \left( \xi_{x} \right) = \max_{\beta,\left(\gamma\ne\beta\right) > 0}\mathcal{K}_{\beta\gamma}^2 \ll 1
\end{align}
Thus, the regime classifier $\chi_{DDC}$ depends on the maximum value of the square of the generalized Knudsen numbers $\mathcal{K}_{\beta\gamma}$ considering all allowed values of the indices $\beta$ and $\gamma\ne\beta$. With this condition satisfied, the GHE [Eq.~\ref{eq:lpbe_frequency_domain_temperature_response}] reduces into an intermediate heat equation (IHE) as: 
\begin{align}
    -i \eta \Delta \tilde{T} + \sum_{\alpha > 0} \frac{\left( \xi_x \mathcal{V}^{ 0 \alpha}_x \right)^{2}}{\Gamma^{\alpha \alpha}_{d}} \Delta \tilde{T} & = \frac{\tilde{h}^{0}}{\zeta}
    \label{eq:intermediate_frequency_domain_temperature_response}
\end{align}
where $\Gamma^{\alpha \alpha}_{d} = -i\eta + \sigma^\alpha + \sum_{\bar{\beta} > 0}\frac{\left(\xi_x\mathcal{V}^{\alpha\bar{\beta}}_{x}\right)\left(\xi_x\mathcal{V}^{\bar{\beta}\alpha}_{x}\right)}{\sigma^{\bar{\beta}}}$ are the elements of $\bm{\Gamma}_d$. As we show in the next sections, Eq.~\ref{eq:intermediate_frequency_domain_temperature_response} can further be reduced into Fourier-diffusive, weakly quasiballistic and hydrodynamic heat equations when additional conditions are imposed on the coefficients of $\Delta\tilde{T}$.

\subsection{Weakly quasiballistic heat flow regime} \label{subsec:weakly_quasiballistic_heat_equation}
To arrive at the weakly quasiballistic heat equation (wQHE, Eq.~\ref{eq:quasiballistic_heat_equation}) from the IHE [Eq.~\ref{eq:intermediate_frequency_domain_temperature_response}], we introduce the weakly quasiballistic classifier condition (wQC, see Appendix~\ref{app_subsec:weakly_quasiballistic_condition_proof}):
\begin{align}
    \chi_{wQC} \left( \eta \right) = & \frac{\eta}{\min \left( \sigma \ne \sigma^0\right)} \ll 1
    \label{eq:fourier_like_condition}
\end{align}
where $\mathcal{K}_t = \frac{\eta}{\min \left( \sigma \ne \sigma^0\right)}$ is the generalized \emph{temporal} Knudsen number. Intuitively, the weakly quasiballistic classifier condition, $\chi_{wQC} \left( \eta \right) \ll 1$, can be understood as follows: the eigenvalues $\sigma^{\alpha\ne 0}$ quantify the rate at which the non-equilibrium components of  $f'_{\lambda}$ in the basis of $\{\mathfrak{e}^\alpha\}$, given by $\vartheta^{\alpha\ne 0}$, decay according to Eq.~\ref{eq:relaxon_lpbe_time_spatial_domain}. Thus, the equilibrium component $\vartheta^0$ remains the only non-zero component of $f'_{\lambda}$ after a timescale $\gtrsim 1/\text{min}\left(\sigma^{\alpha\ne 0}\right)$, and a local equilibrium is established. If this equilibration timescale is much smaller than the relevant heat conduction timescale $\sim 1/\eta$, then the heat flows by a diffusion-dominated mechanism rather than a drifting mechanism, since all collective drifting components of the phonon populations appearing in the eigenmodes of $\bm{\Omega}$ have already decayed within the heat conduction timescale.

When $\chi_{wQC} \left( \eta \right) \ll 1$, as discussed in Appendix~\ref{app_subsec:weakly_quasiballistic_condition_proof}, the IHE [Eq.~\ref{eq:intermediate_frequency_domain_temperature_response}] transforms into:
\begin{align}
    - i \eta \Delta \tilde{T} + \xi_{x}^{2} \left[ \sum_{\alpha > 0} \frac{\left( \mathcal{V}_{x}^{0 \alpha} \right)^{2}}{\sigma^{\alpha}} \mathcal{S}_{\alpha} \right] \Delta \tilde{T} = \frac{\tilde{h}^{0}}{\zeta}
    \label{eq:fourier_like_frequency_domain_temperature_response}
\end{align}
which is the Fourier transform of the wQHE [Eq.~\ref{eq:quasiballistic_heat_equation}], with an impulsive source term on the right hand side. Here, the term in the square bracket represents the suppressed, effective thermal diffusivity ($\rho_{s}$) with $\left( \mathcal{V}_{x}^{0 \alpha} \right)^{2} / \sigma^{\alpha}$ corresponding to the contribution of the $\alpha^{\text{th}}$ eigenmode of $\bm{\Omega}$ to the bulk diffusivity, and $\mathcal{S}_{\alpha} \leq 1$ is the diffusivity suppression function of the corresponding eigenmode due to weakly quasiballistic effects, given by:
\begin{align}
    \mathcal{S}_{\alpha} = \left( 1 + \sum_{\bar{\beta}>0} \frac{\left( \xi_{x} \mathcal{V}_{x}^{\alpha \bar{\beta}} \right)^{2}}{\sigma^{\alpha} \sigma^{\bar{\beta}}} \right)^{-1}.
    \label{eq:eigenmode_suppression_function}
\end{align}
In Ref.~\cite{minnich_phonon_2015}, a similar suppression function was derived for phonons under the relaxation time approximation (RTA) as $\mathcal{S}_{\lambda} = \left( 1 + \mathcal{K}_{\lambda}^{2} \right)^{-1}$, where $\mathcal{K}_{\lambda} = \xi_{x} \Lambda_{\lambda, x}$ is the Knudsen number under the RTA for a phonon mode with mean free path (MFP) $\Lambda_{\lambda, x}$, which has been used to distinguish the weakly quasiballistic regime ($\mathcal{K}_{\lambda}^{2} \sim 1$) from the Fourier heat diffusion regime ($\mathcal{K}_{\lambda}^{2} \ll 1$)~\cite{hua_transport_2014}. In a similar spirit, the suppression function for diffusivity of the eigenmodes of $\bm{\Omega}$ [Eq.~\ref{eq:eigenmode_suppression_function}] can we written as $\mathcal{S}_{\alpha} = 1/(1+\mathcal{K}_{\alpha \alpha}^{2})$, where $\mathcal{K}_{\alpha \alpha} = \xi_x\sqrt{\sum_{\bar{\beta}>0} \frac{\left(\mathcal{V}_{x}^{\alpha \bar{\beta}} \right)^{2}}{\sigma^{\alpha} \sigma^{\bar{\beta}}}}$ is the diagonal part of the generalized spatial Knudsen number defined in Eq.~\ref{eq:knudsen_number_tensor}. This diagonal part of the generalized spatial Knudsen number thus distinguishes the weakly quasiballistic ($\mathcal{K}_{\alpha \alpha}^{2} \sim 1$) from the Fourier-diffusion regime ($\mathcal{K}_{\alpha \alpha}^{2} \ll 1 $) while for both cases, the off-diagonal elements of the generalized spatial Knudsen number remain small ($\mathcal{K}_{\alpha (\beta\ne\alpha)}^2 \ll 1$).

The suppression function for the eigenmodes ($\mathcal{S}_{\alpha}$) derived in this work [Eq.~\ref{eq:eigenmode_suppression_function}], is valid beyond the RTA for the LPBE, which has not been previously established in the literature to the best of our knowledge. Thus, the suppression function $\mathcal{S}_{\alpha}$ will enable the direct experimental extraction of the spectral contribution of the eigenmodes to the overall $\kappa$ using pump-probe-based optical experiments such as the transient grating (TG)~\cite{ravichandran_role_2016}, even in ultrahigh-$\kappa$ materials such as graphene, diamond, boron arsenide and boron nitride where the RTA for the LPBE does not hold~\cite{malviya_failure_2023}. We discuss the application of the theory developed here for the TG in Section~\ref{sec:regime_specific_transient_grating_response} below.

\subsection{Hydrodynamic heat flow regime} \label{subsec:hydrodynamic_heat_equation}
In an ideal limit of momentum-conserving Normal processes (N-processes) dominating phonon collisions, $\bm{\Omega} \approx \bm{\Omega}_N$, where $\bm{\Omega}_N$ is the collision operator for the Normal processes. In this case, apart from $\mathfrak{e}^{0}$, $\bm{\Omega}$ has three (two) other null vectors, corresponding to a drifting equilibrium representing a collective motion of the phonon gas with a single velocity, given by $\phi^{1}_{\lambda,i} = \sqrt{f_{\lambda}^{0} \left( f_{\lambda}^{0} + 1 \right)} \left( \hbar q_{i} \right)/\sqrt{\sum_{\lambda} \left(\hbar q_{i} \right)^{2} f_{\lambda}^{0} \left( f_{\lambda}^{0} + 1 \right)}$ in three (two) dimensional crystals. In this case, Eq.~\ref{eq:intermediate_frequency_domain_temperature_response} can be reduced to a hydrodynamic heat equation describing a pure drifting motion of phonons, as shown in Ref~\cite{hardy_phonon_1970} and the $\kappa$ of the material under steady-state conditions becomes infinite~\cite{guyer_solution_1966, hardy_phonon_1970}. As we deviate from this ideal scenario by introducing weak momentum-dissipative Umklapp processes (U-processes), these drifting eigenmodes tend to have small non-zero, nearly-degenerate eigenvalues, and their contributions to total $\kappa$ become finite but dominate over the contributions from the other eigenmodes of $\bm{\Omega}$~\cite{malviya_efficient_2025, ravichandran_elasticity_2026}.

More generally, to reduce the IHE [Eq.~\ref{eq:intermediate_frequency_domain_temperature_response}] to the HHE [Eq.~\ref{eq:hyperbolic_heat_equation}], we have derived the requirements on the spatial ($\xi_x$) and temporal ($\eta$) frequencies in Appendix~\ref{app_subsec:hydrodynamic_second_sound_condition_proof}. The requirement on $\xi_x$ is:
\begin{align}
    \chi_{HGC} \left( \xi_x \right) = \max_{\alpha > 0} \left( \sum_{\bar{\beta}>0} \frac{ \left( \xi_x \mathcal{V}^{ \alpha \bar{\beta}}_x \right)^{2}}{\sigma^{\bar{\beta}} \sigma^{\alpha}} \right) = \max_{\alpha > 0} \left( \mathcal{K}_{\alpha \alpha} \right)^{2} \ll 1.
    \label{eq:hydrodynamic_geometric_condition}
\end{align}
We refer to this requirement as the hydrodynamic geometry classifier condition (HGC), which imposes an upper limit on $\xi_x$ (or a lower limit on the length scale [$\approx 2\pi/\xi_x$]) for observing hydrodynamic heat flow. The HGC reduces $\Gamma_{d}^{\alpha \alpha} $ in IHE [Eq.~\ref{eq:intermediate_frequency_domain_temperature_response}] to $\approx -i \eta + \sigma^{\alpha}$. To obtain the final form of HHE, the requirement on $\eta$ is (see Appendix~\ref{app_subsec:hydrodynamic_second_sound_condition_proof}):
\begin{align}
    \frac{\sum_{\alpha \in \mathcal{D}} \kappa_{\alpha}}{\sqrt{ \left( \eta/\mu \right)^{2} + 1}} &\gg \sum_{\alpha \notin D} \frac{\kappa_{\alpha}}{\sqrt{ \left( \eta/\sigma^{\alpha} \right)^{2} + 1}}\nonumber\\
    \implies \frac{\sum_{\alpha \in \mathcal{D}} \kappa_{\alpha}}{\sqrt{ \mathcal{K}_t^{2} + 1}} &\gg \sum_{\alpha \notin D} \frac{\kappa_{\alpha}}{\sqrt{ \mathcal{K}_t^2\left(\mu/\sigma^\alpha\right)^{2} + 1}}
    \label{eq:hydrodynamic_temporal_condition}
\end{align}
where, $\mathcal{D}$ is a set of a few eigenmodes of $\bm{\Omega}$ with nearly degenerate eigenvalues ($=\mu$) and $\mathcal{K}_t = \eta/\mu$ is the generalized temporal Knudsen number. This hydrodynamic temporal classifier condition (HTC) [Eq.~\ref{eq:hydrodynamic_temporal_condition}] in the steady-state limit ($\eta \rightarrow 0$), results in a necessary (but not sufficient) condition that depends on the material property only as:
\begin{equation}
    \sum_{\alpha \in \mathcal{D}} \kappa_{\alpha} \gg \sum_{\alpha \notin \mathcal{D}} \kappa_{\alpha}
    \label{eq:necessary_HMC}
\end{equation}
We refer to this \emph{necessary} material requirement as the hydrodynamic material classifier condition (HMC) and the class of materials that satisfy the HMC as \emph{hydrodynamic} materials, while those that do not satisfy HMC as \emph{non-hydrodynamic}. We note that these \emph{drifting} degenerate eigenmodes of $\bm{\Omega}$ belonging to the set $\mathcal{D}$ also have the smallest eigenvalues [i.e., $\mu = \min\left(\sigma \ne \sigma^0\right)$] for hydrodynamic materials that strictly satisfy Eq.~\ref{eq:necessary_HMC}, as we show for graphene at 100 K in the next section. Hence, the definition of the generalized temporal Knudsen numbers as $\mathcal{K}_t = \eta/\mu$ here is consistent with the definition in the previous section.

As detailed in Appendix~\ref{app_subsec:hydrodynamic_second_sound_condition_proof}, the HGC and the HTC in hydrodynamic materials together result in the HHE given by:
\begin{align}
    - \eta^{2} \Delta \tilde{T} - i \mu \eta \Delta \tilde{T} + \xi_{x}^{2} v^{2}_{ud} \Delta \tilde{T} & = \left( -i \eta + \mu \right) \frac{\tilde{h}^{0}}{\zeta}
    \label{eq:second_sound_frequency_domain_temperature_response}
\end{align}
which is the space-time Fourier transform of a damped temperature wave equation~\cite{beardo_observation_2021}, where $v_{ud} = \sqrt{\sum_{\alpha \in \mathcal{D}} \left( \mathcal{V}^{0 \alpha}_{x} \right)^{2}}$ represents the velocity of the undamped temperature wave, with $\mu$ representing the damping coefficient, with its inverse ($1/\mu$) representing the timescale for momentum dissipation to the lattice via the U-processes~\cite{hardy_phonon_1970}. In the limit of complete absence of Umklapp scattering in a material, $\mu = 0$ corresponding to the three (two) drifting null vectors of $\bm{\Omega}_N$ in three (two) dimensional crystals~\cite{pitaevskii_physical_2012}.

The HGC also enforces the diagonal components of the generalized spatial Knudsen numbers $\mathcal{K}_{\alpha\alpha}$ to be much smaller than one ($\mathcal{K}_{\alpha \alpha}^{2} \ll 1$), which, along with the fundamental requirement on the corresponding off-diagonal components ($\mathcal{K}_{\alpha (\beta\ne\alpha)}^{2} \ll 1$), reduces the GHE to HHE for the hydrodynamic materials. Therefore, for the hydrodynamic heat flow regime, all elements of the generalized spatial Knudsen numbers [Eq.~\ref{eq:knudsen_number_tensor}] must be much smaller than one.

\section{Experimental signatures of Fourier-diffusive and non-Fourier heat flow} \label{sec:regime_specific_transient_grating_response}
In this section, we employ the regime classifiers based on the generalized spatial and temporal Knudsen numbers derived in the previous section to predict the experimental conditions for Fourier-diffusive and non-Fourier heat flow regimes in different semiconductors. To this end, we compare the solution of the GHE [Eq.~\ref{eq:lpbe_frequency_domain_temperature_response}] for each material considered here at different temperatures $T_0$ and spatial frequencies $\xi_x$ with those of the regime-specific continuum heat equations --- FHE [Eq.~\ref{eq:fourier_diffusion_heat_equation}], wQHE [Eq.~\ref{eq:quasiballistic_heat_equation}] and HHE [Eq.~\ref{eq:hyperbolic_heat_equation}]. With these comparisons, we can then classify the nature of the predicted temperature dynamics from the GHE as Fourier-diffusive, weakly quasiballistic, hydrodynamic, ballistic or some intermediate transition regime for every pair of $(T_0, \xi_x)$ considered for each material. Since the solution of the GHE represents the solution of the LPBE [Eq.~\ref{eq:phonon_lpbe_time_spatial_domain}] without any approximations, we obtain the former by employing the low-rank framework developed in Ref.~\cite{malviya_efficient_2025} to solve the LPBE in a computationally efficient manner. To predict the spatio-temporal temperature dynamics for each spatial frequency $\xi_x$, we model the initial excitation as $h^{0}\left(x, t\right) = \Delta T_{0} \zeta e^{-i\xi_{x} x} \delta(t)$ and perform a parameter sweep over $\xi_x$.

We note in passing that this set-up resembles that of the transient grating (TG) experiment --- a widely used experimental technique for probing heat flow across regimes ranging from Fourier-diffusive, weakly quasiballistic and hydrodynamic second sound~\cite{johnson_direct_2013, ravichandran_spectrally_2018, huberman_observation_2019, ding_observation_2022}. Since the TG offers a simple configuration involving a single spatial frequency ($\xi_x = k$, where $k = 2\pi/d$ is the TG wave vector and $d$ is the TG period), it allows us to systematically observe the transitions across different heat flow regimes as $k$ is changed for the same sample, making it an ideal choice for testing the predictive power of our framework. Due to this equivalence, we will refer to the computed temperature responses for a spatial wave vector $\xi_x$ as that computed at a TG grating period $d=2\pi/\xi_x$, moving forward.

\begin{figure*}[!ht]
    \centering
    \includegraphics[width=0.85\linewidth]{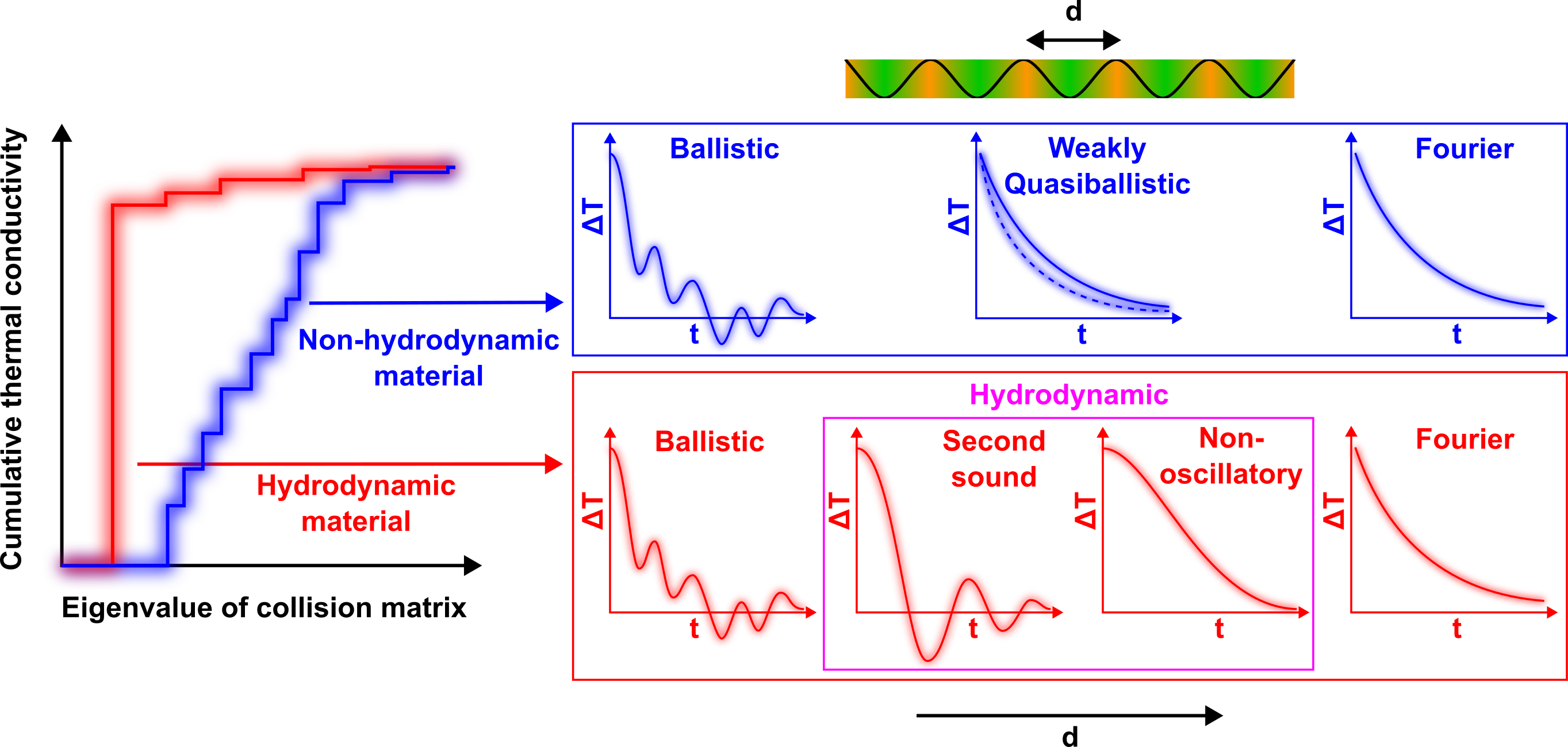}
    \caption{A schematic illustrating the characteristic cumulative $\kappa$ vs. eigenvalues of the collision matrix ($\bm{\Omega}$), along with the transient temperature profiles corresponding to different heat flow regimes observable in the TG experiment for hydrodynamic and non-hydrodynamic materials. Non-hydrodynamic materials show a gradually increasing cumulative $\kappa$ and support weakly quasiballistic heat flow in addition to ballistic and Fourier-diffusive heat flow regimes. On the other hand, hydrodynamic materials exhibit a sharply increasing cumulative $\kappa$ spectrum arising out of large $\kappa$ contributions from a handful of degenerate drifting eigenmodes of $\bm{\Omega}$, thus enabling hydrodynamic heat flow alongside ballistic and Fourier-diffusive heat flow. Within the hydrodynamic regime, we find that the temporal temperature responses can be oscillatory, resulting in a well-known second sound regime of heat transfer, or can be non-oscillatory, non-exponential decay, which has not been reported in the literature to the best of our knowledge.}
    \label{fig:schematic_non_fourier_heat_transfer}
\end{figure*}

In Fig.~\ref{fig:schematic_non_fourier_heat_transfer}, we summarize the heat flow regimes uncovered by the low-rank solution of the LPBE in different materials and for different TG periods $d$. For the hydrodynamic materials, where large contributions to $\kappa$ originate from a handful of the nearly-degenerate eigenmodes of $\bm{\Omega}$, we observe Fourier-diffusive, hydrodynamic and ballistic heat flow regimes as $d$ decreases (and therefore, as the generalized spatial Knudsen numbers increase). Weakly quasiballistic regime is not evident from our simulations for this class of materials, since the hydrodynamic geometry condition ($\chi_{HGC} \ll 1$) is always satisfied whenever the quasiballistic condition ($\chi_{wQC} \ll 1$) is activated, thus resulting in a uniform suppression function of $\mathcal{S}^\alpha \sim 1$ for all eigenmodes of $\bm{\Omega}$. For the non-hydrodynamic materials, we observe Fourier-diffusive, weakly quasiballistic and ballistic heat flow regimes as $d$ decreases. We do not observe any hydrodynamic heat flow for this class of materials since the material requirement [Eq.~\ref{eq:necessary_HMC}] is not satisfied. 

To quantitatively support these predictions, we report the regime classifiers, spectral $\kappa$'s as functions of the eigenvalues of $\bm{\Omega}$ and the solutions of the GHE as well as the continuum field equations for twenty two different group IV and III-V cubic semiconductors across a broad temperature range from 60~K to 100~K in the following section and in the Supplemental Material, sections S1 and S2. We chose these materials, since they have recently attracted significant interest among the experimental research community, owing to the observations of ultra-high $\kappa$ through conventional (e.g., diamond~\cite{inyushkin_thermal_2018, inyushkin_thermal_2025}) as well as unconventional, phonon band structure-engineered routes (e.g., boron arsenide~\cite{kang_experimental_2018, li_high_2018, tian_unusual_2018}), giant $\kappa$-enhancement upon isotopic enrichment in boron nitride~\cite{chen_ultrahigh_2020} and unusually large effect of the isotopic mass of boron on the $\kappa$ of enriched boron phosphide~\cite{zhu_vapor-flux_2026}. Furthermore, the quality of lab-grown and purified samples of these materials has significantly improved in the recent past~\cite{zhu_vapor-flux_2026, chen_ultrahigh_2020}. We find that strong hydrodynamic signatures are challenging to achieve in these materials under laboratory conditions due to the requirements of ultra-low temperatures and large samples of isotope-free pristine quality predicted by our calculations. Nevertheless, weakly quasiballistic, ballistic and Fourier-diffusive regimes along with a transition from weakly hydrodynamic to ballistic regimes are observable in several of these materials in the range of temperatures considered here. Motivated by the recent prediction of strong hydrodynamic signatures in graphene at modest temperatures of 150~K~\cite{ravichandran_elasticity_2026, malviya_indicators_2026}, we have also solved the GHE at 100~K for different TG periods in this material, and have uncovered strong signatures of Fourier-diffusive, hydrodynamic and ballistic heat flow regimes under experimentally accessible conditions. The first principles method used to obtain these results is discussed in Appendix~\ref{app_sec:first_principles_desciprion} and in the Supplemental Material, section S4.

\begin{figure}[!ht]
    \centering
    \includegraphics[width=0.8\linewidth]{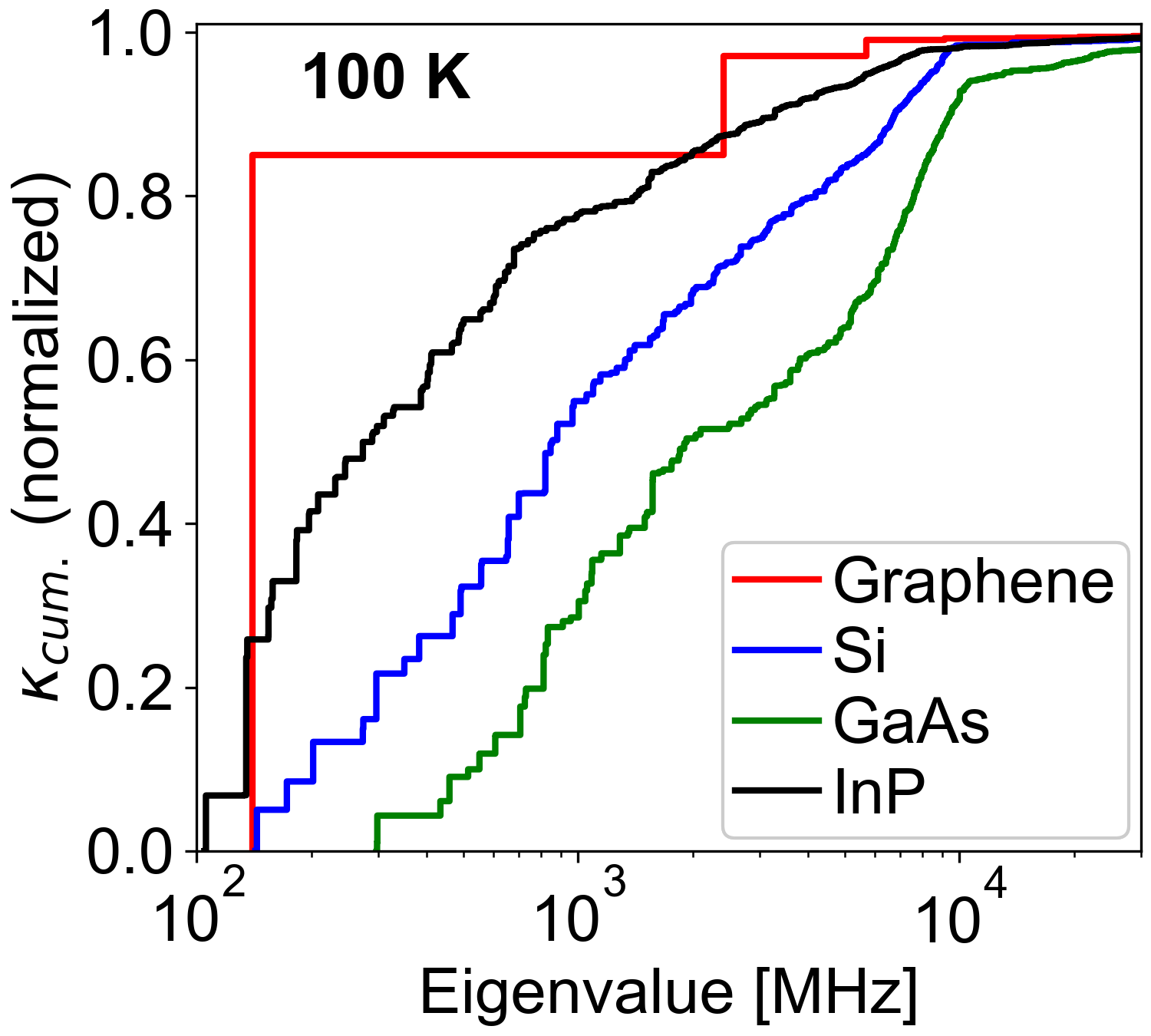}
    \caption{Normalized cumulative $\kappa$ vs. eigenvalues of the collision matrix arranged in an increasing order for graphene, silicon (Si), gallium arsenide (GaAs), indium phosphide (InP) at 100~K. Among these materials, graphene is a hydrodynamic material, while the remaining materials are non-hydrodynamic at 100~K, based on the distinction illustrated in Fig.~\ref{fig:schematic_non_fourier_heat_transfer}.}
    \label{fig:kappa_spectrum_GaAs_InP_Si_Graphene}
\end{figure}

We begin the discussion with the cumulative $\kappa$ [$\kappa_{cum.}$] from the eigenmodes of $\bm{\Omega}$ with increasing eigenvalues for graphene, Si, GaAs, and InP at 100~K in Fig.~\ref{fig:kappa_spectrum_GaAs_InP_Si_Graphene}. We are able to predict the Fourier-diffusive as well as all non-Fourier heat flow regimes introduced earlier in these four materials at 100~K, so we have presented the supporting results for other III-V and group-IV compounds in the Supplemental Material, sections S1-S3. For graphene, we observe $\sim$ 85\% contribution to $\kappa$ from two drifting eigenmodes of $\bm{\Omega}$ associated with the smallest eigenvalue, indicating that it can be classified as a hydrodynamic material at 100~K. However, for other materials shown in Fig.~\ref{fig:kappa_spectrum_GaAs_InP_Si_Graphene}, the $\kappa_{cum.}$ gradually increases with increasing eigenvalues of $\bm{\Omega}$ and lacks any large contribution from individual eigenmodes, indicating that these are non-hydrodynamic materials.

\subsection{Weakly quasiballistic regime} \label{subsec:transient_grating_weakly_quasiballistic}
In the weakly quasiballistic heat flow regime (i.e., when $\chi_{DDC} \ll 1$ and $\chi_{wQC} \ll 1$), the temperature evolves according to the wQHE [Eq.~\ref{eq:fourier_like_frequency_domain_temperature_response}]. In this case, the temperature response ($\Delta T$), obtained by calculating the inverse Fourier transform of $\Delta \tilde{T}$, is given by:
\begin{align}
    \Delta T \left(x, t \right) = \Delta T_{0} e^{-ikx} e^{-\rho_{s} k^{2} t} u \left(t \right)
    \label{eq:fourier_like_TG_time_domain_temperature_response}
\end{align}
where $u \left(t \right)$ is a unit step function and $\rho_{s} k^{2}$ is the decay rate suppressed due to the weakly quasiballistic effect. As discussed in the previous section (section~\ref{sec:generalized_regime_classifiers}), a significant suppression due to a weakly quasiballistic effect requires $\mathcal{K}_{\alpha \alpha}^{2} \sim 1$ at least for one eigenmode $\alpha$ ($\equiv \chi_{HGC} \sim 1$). For $\chi_{HGC} \ll 1$, the largest $\mathcal{K}_{\alpha \alpha}^{2} \ll 1$ and the heat flow transitions into the Fourier-diffusive regime. 

\begin{figure}[!ht]
    \centering
    \includegraphics[width=1.0\linewidth]{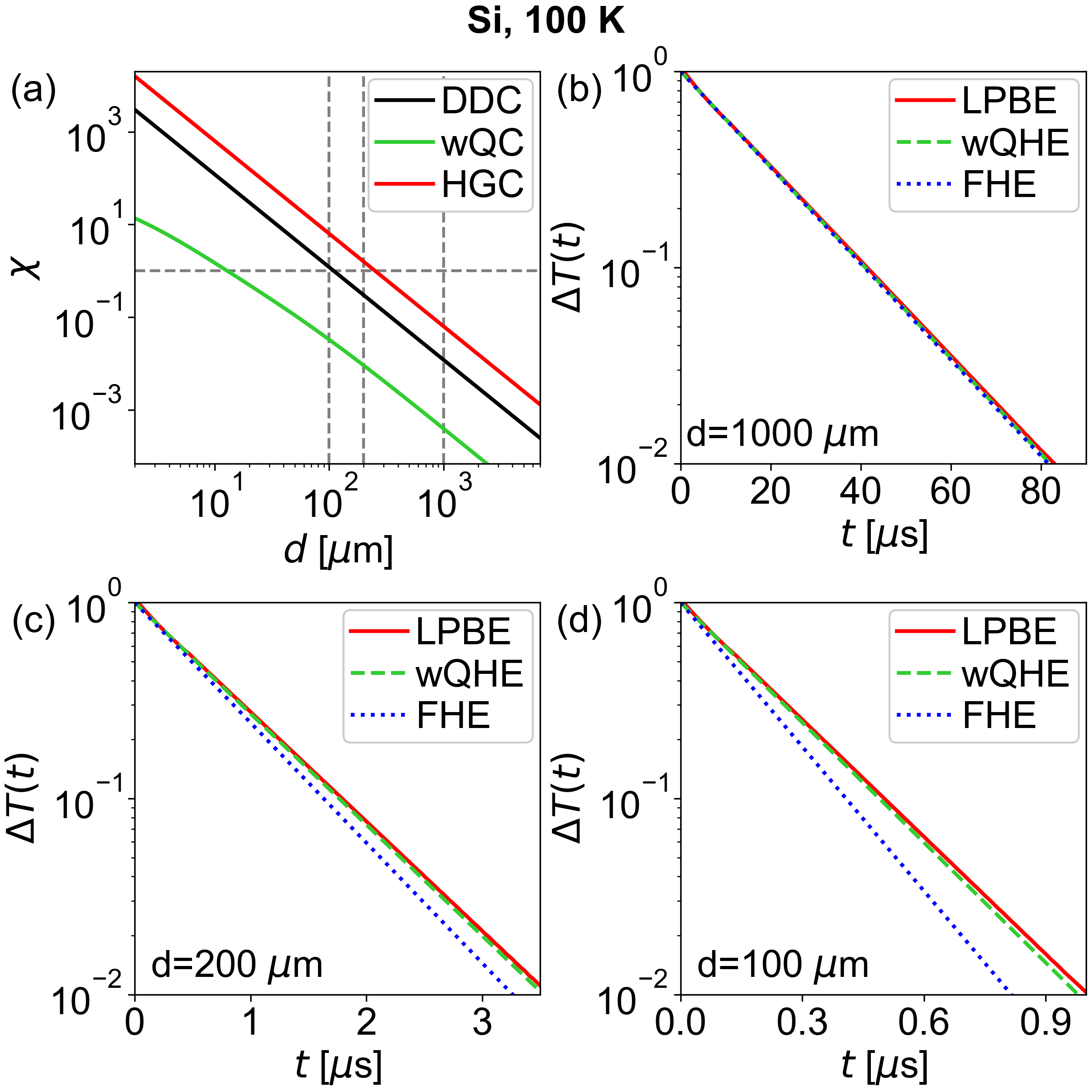}
    \caption{Regime classifiers $\chi$ and the predicted transient temperature responses at three grating periods for Si at 100~K. (a) Variation of the regime classifiers from the diagonal dominance condition ($\chi_{DDC}$), the weakly quasiballistic condition ($\chi_{wQC}$) and the hydrodynamic geometry condition ($\chi_{HGC}$) as functions of the grating period ($d$). Figs.~(b)-(d) show the temperature evolution for three different grating periods obtained by solving the LPBE, the wQHE and the FHE. Here, Fig.~(b) shows the Fourier-diffusive heat flow regime observed at $d=1000\ \mu$m, where the LPBE solution overlaps with the solution of the FHE. Fig.~(c) shows the weakly quasiballistic heat flow regime observed at $d=200\ \mu$m, where the LPBE solution decays more slowly than that from the FHE, while it matches with the solution of the wQHE. As $\chi_{DDC}$ approaches 1 [e.g., for Fig.~(d) at $d=100\ \mu$m], the LPBE solution shows deviation from the wQHE solution as well, indicating the onset of the transition into a ballistic heat flow regime. The grating periods for Figs.~(b)-(d) are marked by dashed vertical lines in the plots for regime classifiers [Fig.~(a)]. Here, the LPBE solutions are obtained using the low-rank method from Ref.~\cite{malviya_efficient_2025}, where we used $\sim$36\% of the eigenmodes, corresponding to 99\% of the total $\kappa$ for Si at 100~K.}
    \label{fig:regime_specific_condition_tg_soln_Si}
\end{figure}

For such an exponentially decaying temperature response, the relevant temporal frequency $\eta$ in wQC [Eq.~\ref{eq:fourier_like_condition}] becomes comparable to the thermal decay rate $\rho_{s} k^{2}$. Therefore the wQC can be written in terms of the grating wave vector $k$ as $ \chi_{wQC} \left( k \right) = \rho_{s} k^{2} /\min \left( \sigma \ne \sigma^0\right) \ll 1$. In Fig.~\ref{fig:regime_specific_condition_tg_soln_Si}~(a), we show the regime classifier conditions for different heat flow regimes as we vary the grating period $d$ for Si at 100~K. Here, for grating periods longer than 100~$\mu$m, both DDC as well as wQC are activated ($\chi_{DDC} \ll 1$ and $\chi_{wQC} \ll 1$), and so, we predict weakly quasiballistic heat flow in this region. Further, as $d$ increases, the weakly quasiballistic effect diminishes since $\chi_{HGC}$ --- a measure of the suppression of thermal diffusivity due to the weakly quasiballistic effect --- also reduces, leading to Fourier-diffusive heat flow, as discussed in Section~\ref{subsec:weakly_quasiballistic_heat_equation}.

We validate our predictions by calculating the temporal temperature evolution for three different grating periods: $d = $ 1000~$\mu$m, 200~$\mu$m, 100~$\mu$m for Si at 100~K, as shown in Figs.~\ref{fig:regime_specific_condition_tg_soln_Si}~(b), (c) and~(d), respectively. For $d=$~1000~$\mu$m, the solutions of the wQHE and the FHE closely match the full LPBE solution, thus confirming Fourier-diffusive heat flow with bulk $\kappa$ under these conditions. For $d=$~200~$\mu$m, the LPBE solution agrees well with the wQHE solution, but deviates from the FHE solution, indicating a weakly quasiballistic heat flow regime. For both of these grating periods, the diagonal dominance condition ($\chi_{DDC} \ll 1$) remains active. However, as $d$ is reduced further, $\chi_{DCC}$ approaches 1, and so, the reduction of the LPBE to wQHE is no longer formally possible. In Fig.~\ref{fig:regime_specific_condition_tg_soln_Si} (d), we find that the solution of LPBE starts deviating from the solution of wQHE for $d=$~100~$\mu$m indicating an onset of regime transition to ballistic heat flow. Nevertheless, the observed differences between the solutions of the LPBE and the wQHE are small around this transition regime ($\chi_{DDC} \sim 1$), and may not be discernible in the experiments due to the measurement noise. 

\begin{figure}[!ht]
    \centering
    \includegraphics[width=1.0\linewidth]{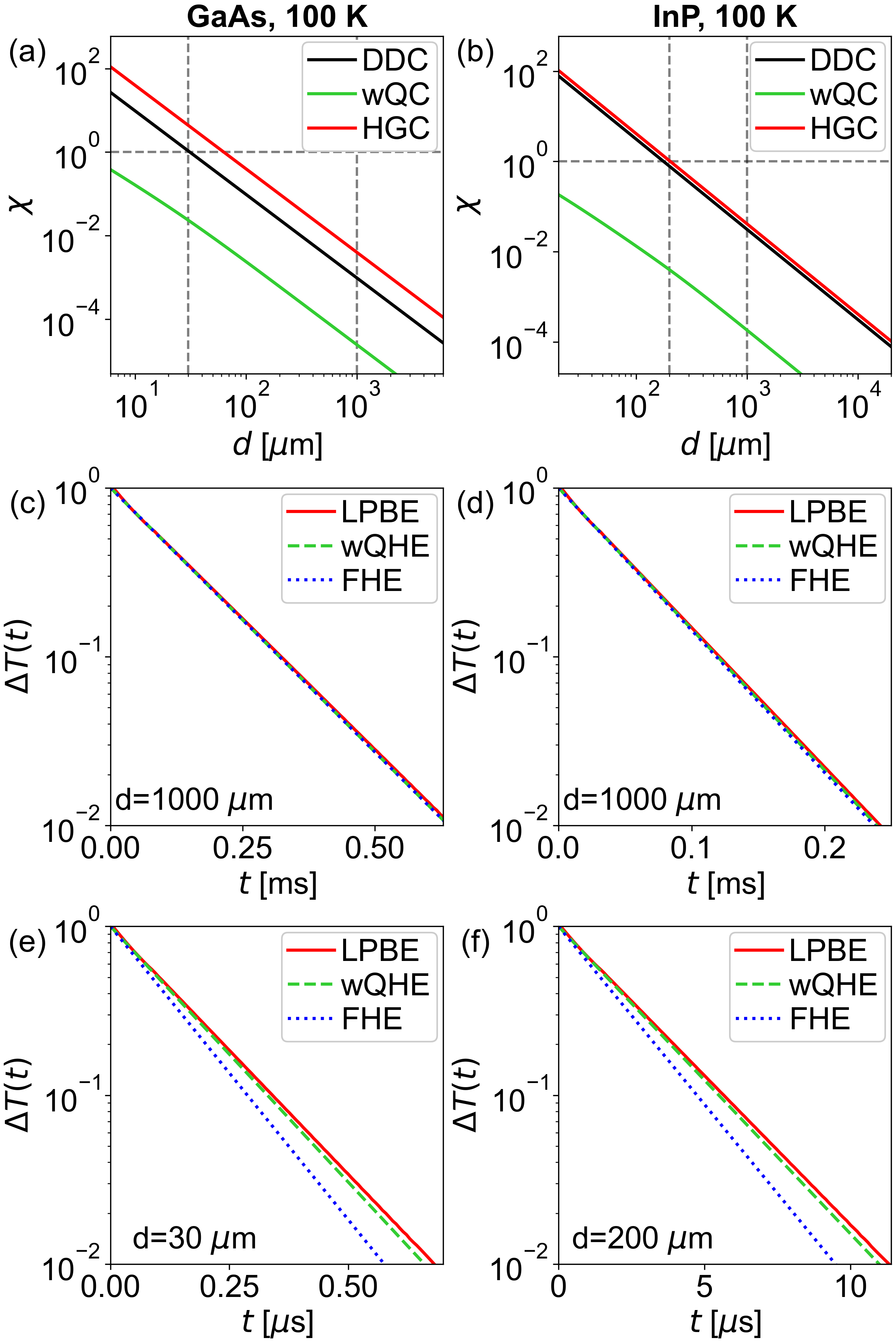}
    \caption{Regime classifiers [$\chi_{DDC}$, $\chi_{wQC}$ and $\chi_{HGC}$] and transient temperature responses at two grating periods for GaAs (1st column), and InP (2nd column) at 100~K. Similar to Fig.~\ref{fig:regime_specific_condition_tg_soln_Si}, Figs.~(a) and~(b) show the variation of regime classifiers ($\chi$) vs. the grating period ($d$). Figs.~(c)-(f) show the temperature evolution in these materials for two different grating periods, with Figs.~(c) and~(d) representing the Fourier-diffusive regime and Figs.~(e) and~(f) representing the transition from the weakly quasiballistic to the ballistic regime. These grating periods are marked by dashed vertical lines in the plots for regime classifiers [Figs.~(a) and~(b)]. Here, the LPBE solutions are obtained using the low-rank method from Ref.~\cite{malviya_efficient_2025}, where we used $\sim$54\% and 70\% of the eigenmodes, corresponding to 99\% of the total $\kappa$, for GaAs and InP, respectively at 100~K.}
    \label{fig:regime_specific_condition_tg_soln_GaAs_InP}
\end{figure}

In Fig.~\ref{fig:regime_specific_condition_tg_soln_GaAs_InP}, we have shown similar results for GaAs and InP at 100~K. For these materials, our regime classifier conditions predict weakly quasiballistic and Fourier regimes for $d$ greater than 30~$\mu$m and 200~$\mu$m, respectively, as shown in Fig.~\ref{fig:regime_specific_condition_tg_soln_GaAs_InP}~(a) and~(b). Corresponding temperature responses are shown in Figs.~\ref{fig:regime_specific_condition_tg_soln_GaAs_InP}~(c)-(f), which exhibit Fourier-diffusive heat flow for $d=$~1000~$\mu$m in both GaAs and InP, and a regime transition from weakly quasiballistic heat flow with maximum suppression to ballistic heat flow at $\chi_{DDC} \sim 1$ for $d=$~30~$\mu$m and 200~$\mu$m for GaAs and InP, respectively. These results demonstrate that our regime classifiers not only correctly identify the heat flow regimes but also provide an analytical solution of LPBE (by identifying the appropriate continuum equations and their parameters to employ) without the need for a complete numerical solution that can be computationally intensive for such non-hydrodynamic materials~\cite{malviya_efficient_2025}.

\begin{figure}[!ht]
    \centering
    \includegraphics[width=0.8\linewidth]{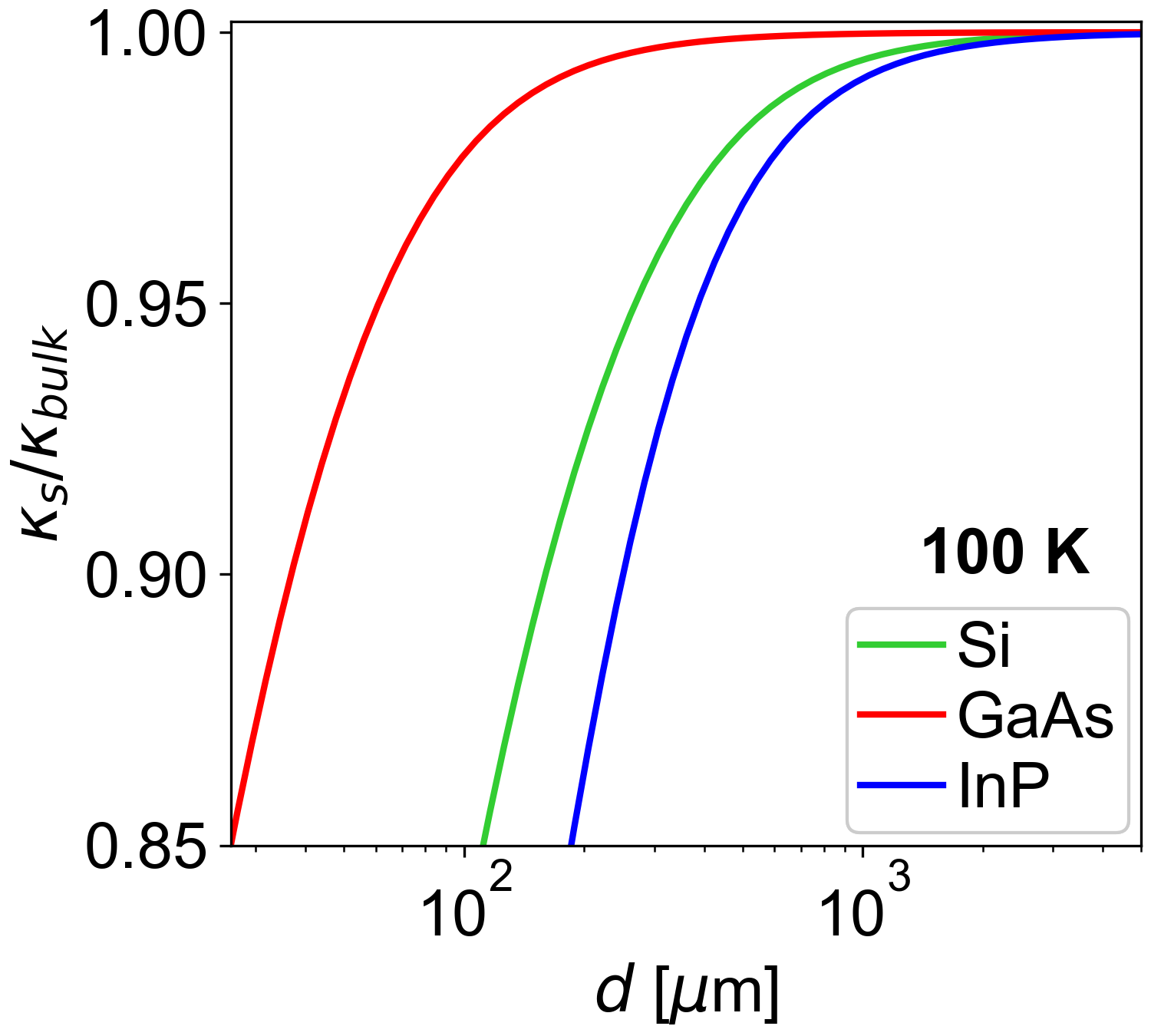}
    \caption{Ratio of the apparent, suppressed thermal conductivity ($\kappa_{s}$) to its bulk counterpart ($\kappa_{bulk}$) as a function of grating period ($d$), for Si, GaAs, and InP at 100~K. As $d$ decreases, the suppression of $\kappa$ becomes stronger, leading to more pronounced weakly quasiballistic heat flow.}
    \label{fig:suppressed_kappa_GaAs_InP_Si}
\end{figure}

In Fig.~\ref{fig:suppressed_kappa_GaAs_InP_Si}, we show the suppression of $\kappa$, compared to its bulk counterpart ($\kappa_{s}/\kappa_{bulk} = \rho_{s}/\rho_{bulk}$, where subscripts \emph{s} and \emph{bulk} represent the suppressed and bulk values, respectively) as a function of $d$, for Si, GaAs, and InP at 100~K. For maximum weakly quasiballistic effect as shown in Figs.~\ref{fig:regime_specific_condition_tg_soln_Si}~(d), ~\ref{fig:regime_specific_condition_tg_soln_GaAs_InP}~(e) and~\ref{fig:regime_specific_condition_tg_soln_GaAs_InP}~(f), where $\chi_{DDC} \lesssim 1$, we see $\sim$15\% reduction in $\kappa$ from its bulk value for all three materials at 100~K, as shown in Fig.~\ref{fig:suppressed_kappa_GaAs_InP_Si}.

\subsection{Hydrodynamic regime} \label{subsec:transient_grating_hydrodynamic_second_sound}
In the hydrodynamic heat flow regime, the temperature in a TG experiment at a grating wave vector $k$ evolves following Eq.~\ref{eq:second_sound_frequency_domain_temperature_response}, whose solution in the frequency domain is given by:
\begin{align}
    \Delta \tilde{T} \left( \xi_{x}, \eta \right) = 2 \pi \Delta T_{0} \frac{\left( -i \eta + \mu \right) \delta \left(\xi_{x} - k \right)}{- \eta^{2} - i \mu \eta + \xi_{x}^{2} v^{2}_{ud}}
    \label{eq:second_sound_TG_frequency_domain_temperature_response}
\end{align}
and upon applying the inverse Fourier transform in space and time, we get the space-time dependent temperature response as:
\begin{align}
    \Delta T \left( x, t \right) & = \Delta T_{0} e^{-ikx} e^{- \mu t / 2} u \left( t \right) \nonumber \\
    \times & \left[ \frac{\mu}{\sqrt{D}} \sin{\left( \frac{\sqrt{D}}{2}t \right)} + \cos{\left( \frac{\sqrt{D}}{2}t \right)} \right]
    \label{eq:second_sound_TG_time_domain_temperature_response}
\end{align}
where $D = 4 v_{ud}^{2} k^{2} - \mu^{2}$ and the relaxation time ($1/e$ time) for the exponentially decaying component is $2/\mu$.

We identify several important features of the hydrodynamic heat flow regime from Eq.~\ref{eq:second_sound_TG_time_domain_temperature_response}. 
\begin{enumerate}
    \item When $D > 0$, the term within the square brackets of Eq.~\ref{eq:second_sound_TG_time_domain_temperature_response} evolves in time as an oscillatory function, while for $D \leq 0$, it ceases to show any oscillatory behavior, and we obtain a monotonically decaying response, as discussed later in this subsection. Thus, the \emph{necessary} (but not sufficient) condition for the HHE [Eq.~\ref{eq:second_sound_frequency_domain_temperature_response}] to admit a damped oscillatory temperature response is:
\begin{align}
    \chi_{OC} = \left[ \frac{\mu}{2 v_{ud} k} \right]^{2} < 1   \label{eq:OC_v1}
\end{align}
which we call the oscillatory classifier condition (OC) within the hydrodynamic regime.
\begin{figure}[!ht]
    \centering
    \includegraphics[width=0.85\linewidth]{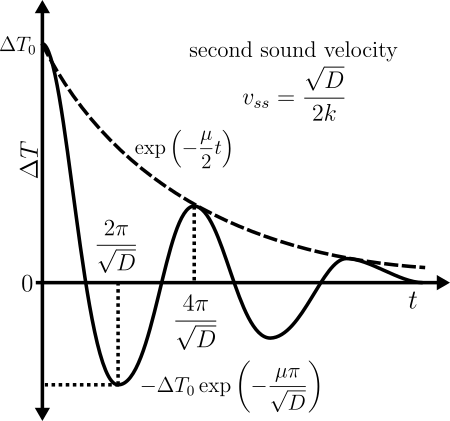}
    \caption{A schematic illustrating the oscillatory temperature response given by Eq.~\ref{eq:second_sound_TG_time_domain_temperature_response}, which represents the temporal evolution of temperature in the transient grating experiment and at a spatial location $x = 0$. This figure illustrates the key features of the oscillations, including the periodicity and the resulting second sound velocity ($v_{ss}$), the depth of the first dip, and the blanket exponential envelope, as detailed in the main text.}
    \label{fig:bte_to_hyperbolic_equation}
\end{figure}

Figure~\ref{fig:bte_to_hyperbolic_equation} shows a schematic of the temporal solution of the HHE [Eq.~\ref{eq:second_sound_TG_time_domain_temperature_response}] at a spatial location $x=0$ when $\chi_{OC} < 1$ is satisfied. Here, with a spatially-sinusoidal instantaneous initial temperature distribution, diffusive thermalization appears as an exponential temporal decay of the temperature at $x=0$~\cite{johnson_direct_2013, ravichandran_spectrally_2018}, while an advective (but damped) temperature wave appears as an oscillatory temporal decay, with transient peaks appearing in the temperature response, when the adjacent advective crests of the initial sinusoidal temperature profiles cross each other at $x=0$. This non-monotonic oscillatory temperature profile is the classical signature of the second sound regime in the TG experiment~\cite{huberman_observation_2019, ding_observation_2022}.

\item Though $D>0$ (or equivalently, $\chi_{OC} < 1$) ensures that the terms within the square brackets in Eq.~\ref{eq:second_sound_TG_time_domain_temperature_response} are oscillatory, these oscillations are visible only when the oscillation frequency is significantly more than the decay rate of the background exponential ($\mu/2$); otherwise the background exponential suppresses the oscillatory component of the temporal temperature response before completing one complete cycle and so, appears to decay towards zero from below the time axis. The requirement for complete oscillations to be visible is directly obtained from the features of the frequency domain solution [$\Delta \tilde{T}\left(x=0, \eta\right)$], where a complete oscillatory response will manifest as a broadened spike at a non-zero frequency in $| \Delta \tilde{T} \left( \eta \right)|$, as shown in Fig.~\ref{fig:schematic_fourier_analysis}~(b) in the Appendix section~\ref{app_sec:frequency_analysis}. This requirement further restricts the upper limit of $\chi_{OC}$ to:
\begin{equation}
    \chi_{OC} < \left( \sqrt{2}+1 \right) /4 \approx 0.6. 
    \label{eq:visibility_condition}
\end{equation}

\item In the second sound regime, the velocity of the advective temperature wave is proportional to the ratio of the grating period to the time between adjacent positive peaks in the temporal temperature response. From Fig.~\ref{fig:bte_to_hyperbolic_equation}, this velocity, which is the second sound velocity ($v_{ss}$), is given by $v_{ss} = \sqrt{D}/2k$, and the depth of the first dip is given by $\Delta T_{0} \exp ( - \mu \pi / \sqrt{D} )$. Interestingly, both of these quantities, which have been used as signatures of the second sound regime in the past computational~\cite{ding_umklapp_2018, zhang_emergence_2022} as well as experimental~\cite{huberman_observation_2019, ding_observation_2022} works, are explicitly dependent on the wave vector $k$, and therefore, the length scale of temperature gradients in the experiment. Only in the limit of purely hydrodynamic heat flow, where the pure drift eigenmodes of the collision matrix of the N-processes ($\bm{\Omega}_N$) with $\mu = 0$ are the dominant heat carriers, the second sound velocity and the depth of the oscillatory temperature responses become independent of the heating length scale.

\item For $D \le 0$ (or $\chi_{OC} \ge 1$), the temperature evolves as a non-oscillatory, non-exponential (therefore not weakly quasiballistic), monotonically decreasing function of $t$ given by:
\begin{align}
    \Delta T \left( x, t \right) & = \Delta T_{0} e^{-ikx} e^{- \mu t / 2} u \left( t \right) \nonumber \\
    \times & \left[ \frac{\mu}{\sqrt{D'}} \sinh{\left( \frac{\sqrt{D'}}{2}t \right)} + \cosh{\left( \frac{\sqrt{D'}}{2}t \right)} \right]
    \label{eq:non_oscilatory_second_sound_TG_time_domain_temperature_response}
\end{align}
where $D' = -D$. This new non-oscillatory feature of hydrodynamic heat flow regime has not been reported in the literature to the best of our knowledge. Thus, even when the HHE describes the heat flow in a material, a second sound signature is not necessarily guaranteed. We further note that, this new variant of the hydrodynamic regime cannot occur in the limit of ideal hydrodynamic heat flow where the Umklapp processes are completely absent and so, $\mu = 0$, since $D$ will always be positive in this limit.

\item For long grating periods ($k \rightarrow 0$), Eq.~\ref{eq:non_oscilatory_second_sound_TG_time_domain_temperature_response} reduces to an exponential solution as in the Fourier-diffusive regime, with temporal decay rate $v_{ud}^{2} k^{2} / \mu$ (see Appendix~\ref{app_sec:hha_at_large_grating}). Since $v_{ud}^{2}/\mu$ is the thermal diffusivity from the eigenmodes of $\bm{\Omega}$ belonging to the special set $\mathcal{D}$ in hydrodynamic materials introduced earlier,  Eq.~\ref{eq:non_oscilatory_second_sound_TG_time_domain_temperature_response} reduces exactly to the solution of FHE, with the contribution to the diffusivity originating from the eigenmodes belonging to the set $\mathcal{D}$. For such large $d$, the deviation in Eq.~\ref{eq:non_oscilatory_second_sound_TG_time_domain_temperature_response} from the solution of the FHE will be smaller when the strength of the hydrodynamic nature of the material is larger, since the contribution to the thermal diffusivity from the non-drifting eigenmodes of $\bm{\Omega}$, which appears in the FHE but not in the HHE, is smaller.
\end{enumerate}

\begin{figure}[!ht]
    \centering
    \includegraphics[width=0.85\linewidth]{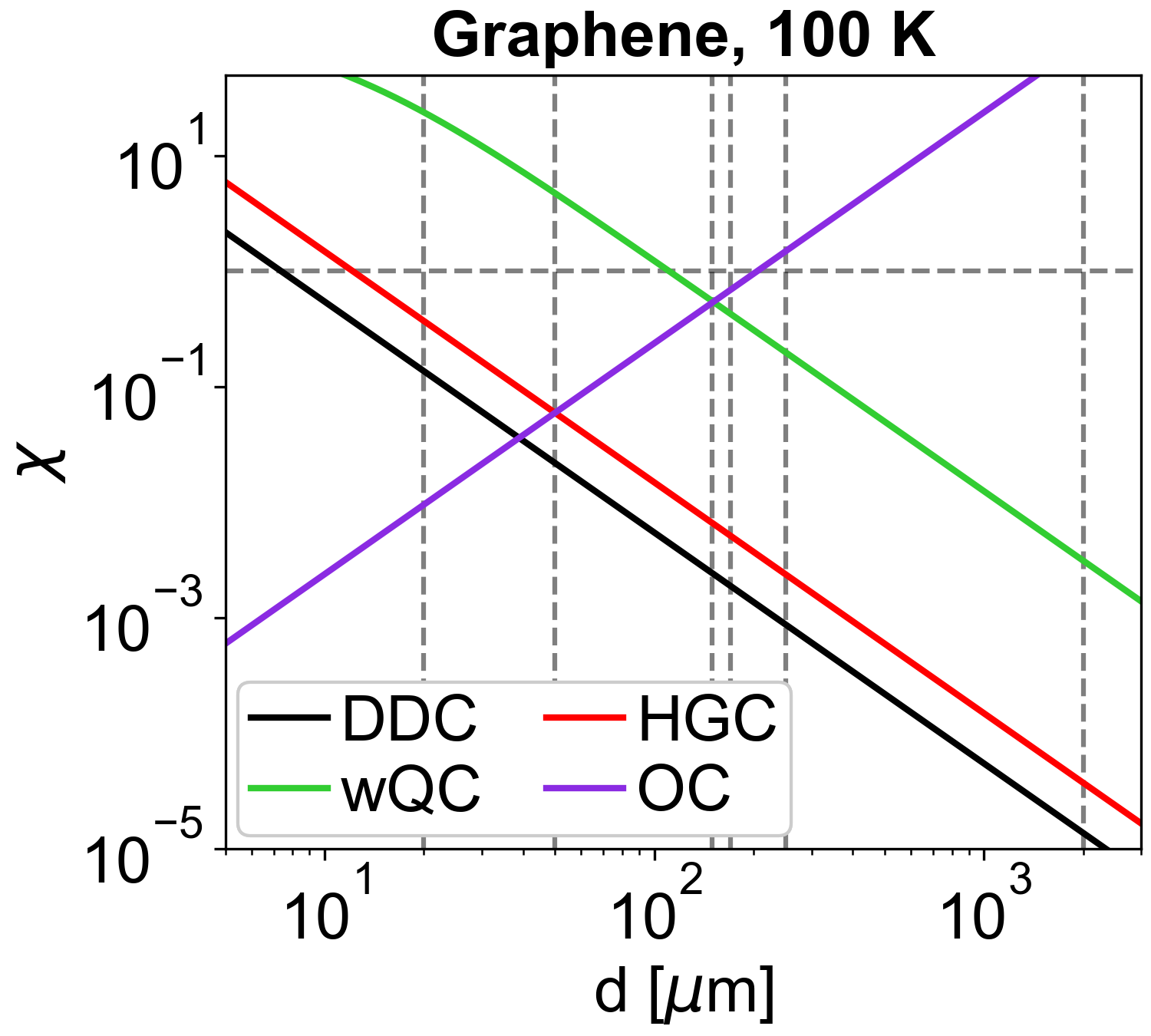}
    \caption{Regime classifiers [$\chi_{DDC}$, $\chi_{wQC}$, $\chi_{HGC}$ and $\chi_{OC}$] vs. grating period ($d$) for graphene at 100~K. Conditions for hydrodynamic second sound are activated for $d \gtrsim$~15~$\mu$m, but will not exhibit oscillations beyond $d \sim$~160~$\mu$m due to the deactivation of OC ($\chi_{OC} \gtrsim 0.6$). For longer $d$, a non-oscillatory hydrodynamic regime emerges and finally transitions into a Fourier-diffusive regime. The transient temperature response demonstrating different regimes are shown in Fig.~\ref{fig:tg_soln_Graphene} for the grating periods are marked by dashed vertical lines.}
    \label{fig:regime_specific_condition_Graphene}
\end{figure}

To demonstrate these features of hydrodynamic heat flow, we perform calculations for graphene at 100~K, which satisfies the hydrodynamic material condition as shown in Fig.~\ref{fig:kappa_spectrum_GaAs_InP_Si_Graphene}, with $\mu \sim$~140~MHz, and $v_{ud} \sim$~ 2300~m/s, and the corresponding regime classifiers are shown in Fig.~\ref{fig:regime_specific_condition_Graphene}. From this figure, we predict the hydrodynamic second sound temperature response for $d\gtrsim$~15~$\mu$m, where the LPBE reduces to the HHE. However, the oscillatory response will be visible only until $d \sim$~160~$\mu$m where $\chi_{OC} \lesssim 0.6$, beyond which, complete oscillations will no longer be visible. For $d \gtrsim$~200~$\mu$m ($\chi_{OC} > 1$), the temperature response will transition into a monotonic, non-oscillatory, non-exponential hydrodynamic decay [Eq~\ref{eq:non_oscilatory_second_sound_TG_time_domain_temperature_response}], following which a gradually transition to the Fourier-diffusive regime would occur for $d \gg$~200~$\mu$m.

\begin{figure}[!ht]
    \centering
    \includegraphics[width=1.0\linewidth]{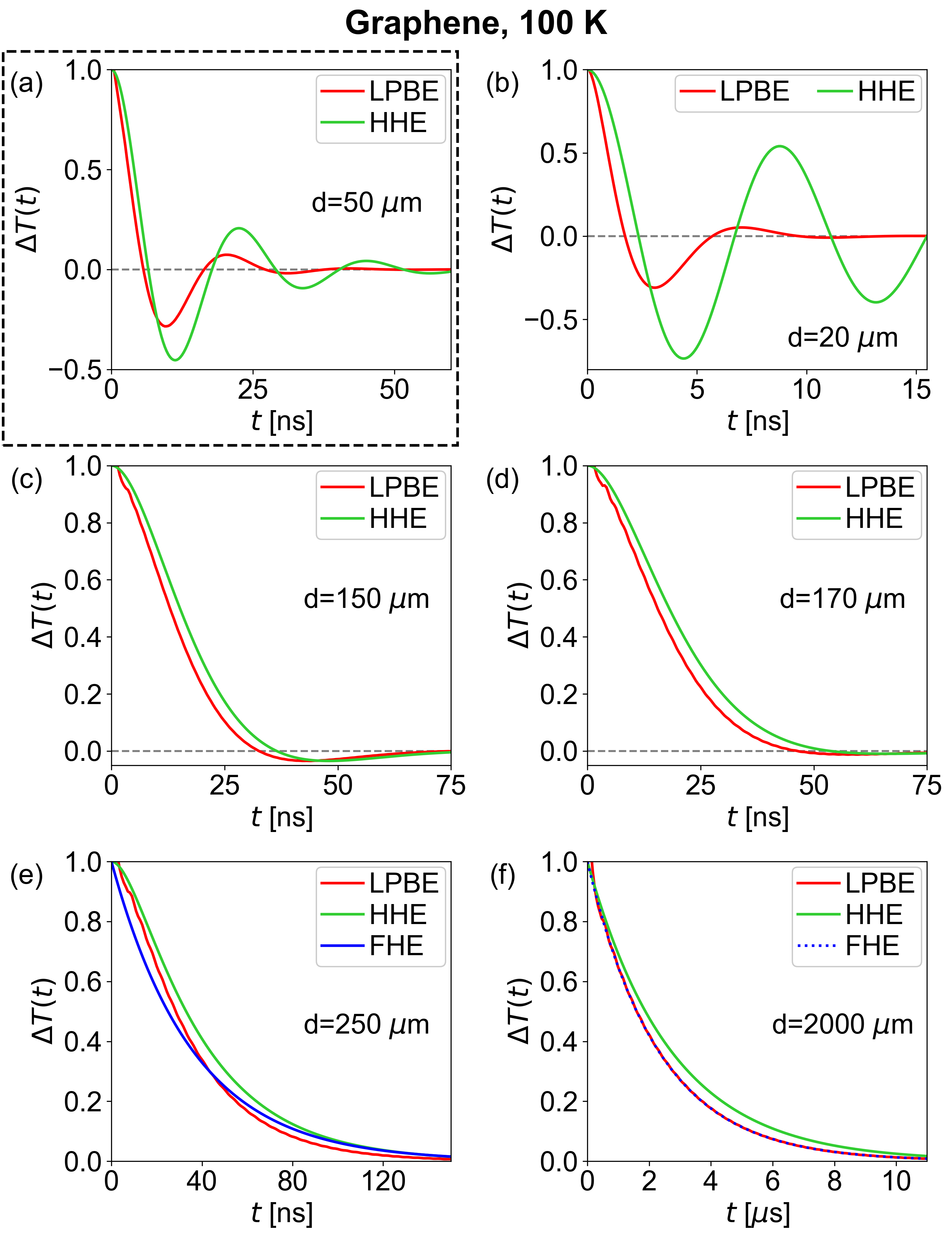}
    \caption{The transient temperature responses for graphene at 100~K and grating periods of (a) 50~$\mu$m, (b) 20~$\mu$m, (c) 150~$\mu$m, (d) 170~$\mu$m, (e) 250~$\mu$m, and (f) 2000~$\mu$m, obtained using the full solutions of the LPBE as well as the solutions of the HHE, and the FHE. We observe hydrodynamic second sound with complete oscillations at the grating periods of 50~$\mu$m,  20~$\mu$m and  150~$\mu$m. At a grating period of 170~$\mu$m, the oscillation dies down before completing a full cycle and the transient response appears to approach zero from below the time axis. For 250~$\mu$m, we observe a new type of non-oscillatory transient hydrodynamic response, that appears as a transition regime from an oscillatory second sound response at smaller $d$ to a Fourier-diffusive regime at longer $d$. Here, the LPBE solutions are obtained using the low-rank method from Ref.~\cite{malviya_efficient_2025}, where we used $\sim$9\% the eigenmodes, corresponding to 99.7\% of the total $\kappa$ for graphene at 100~K.}
    \label{fig:tg_soln_Graphene}
\end{figure}

To validate these predictions, in Fig.~\ref{fig:tg_soln_Graphene}, we show the transient temperature responses for different grating periods corresponding to different heat flow regimes. We observe hydrodynamic second sound with clearly visible oscillations and strong periodic dips in the temperature response at 50~$\mu$m, as shown in Fig.~\ref{fig:tg_soln_Graphene}~(a). The grating period of 50~$\mu$m corresponds to an intersection of two opposing regime classifiers, $\chi_{HGC}$ and $\chi_{OC}$ (see Fig.~\ref{fig:regime_specific_condition_Graphene}). To the left of this point ($d < $~50~$\mu$m), $\chi_{OC}$ is further reduced; therefore, we predict the oscillations with a larger dip. However, at the same time, $\chi_{HGC}$ is larger, which weakens the validity of reducing the LPBE into the HHE, resulting in a larger deviation from the ideal second sound, as shown in Fig.~\ref{fig:tg_soln_Graphene}~(b) for 20~$\mu$m grating period. On the other hand, for $d > $~50~$\mu$m, $\chi_{OC}$ increases and $\chi_{HGC}$ reduces, resulting in a smaller deviation of the LPBE solution from the HHE solution albeit with a reduced amplitude of the temperature dip, as shown in Fig.~\ref{fig:tg_soln_Graphene}~(c) for 150~$\mu$m grating period. Fig.~\ref{fig:tg_soln_Graphene}~(d) shows such a transient temperature response with barely visible oscillations at long times (50-75~ns) for a 170~$\mu$m grating period, where $\chi_{OC} \sim 0.7$.

Further, for $d=$~250~$\mu$m in Fig.~\ref{fig:tg_soln_Graphene}~(e), we observe a new type of transient hydrodynamic response which is non-oscillatory and non-exponential in nature. This grating period lies in the transition region between the oscillatory second sound regime and the Fourier-diffusive regime. The non-exponential nature of the corresponding transient temperature response is established by comparing with the solution of the FHE for this grating period in Fig.~\ref{fig:tg_soln_Graphene}~(e). Finally, for a 2000~$\mu$m grating period, we observe Fourier-diffusive heat flow, where the FHE solution overlaps with the LPBE solution as shown in Fig.~\ref{fig:tg_soln_Graphene}~(f). In all of these cases, the HHE solutions agree qualitatively but deviate quantitatively from the full solutions of LPBE, since the degenerate drifting eigenmodes of $\bm{\Omega}$ in graphene at 100~K contribute to $\sim$85\% of the total $\kappa$ (see Fig.~\ref{fig:kappa_spectrum_GaAs_InP_Si_Graphene}), while the remaining eigenmodes are diffusely contributing to heat flow. In fact, as a result of this additional diffuse contribution to $\kappa$ in graphene at 100~K, the solution of the HHE, which contains the contributions to the thermal diffusivity from the drifting eigenmodes only, decays at a $~\sim 15\%$ slower rate compared to those of the FHE and the LPBE for a grating period of 2000 $\mu$m in Fig.~\ref{fig:tg_soln_Graphene}~(f).

\begin{figure}[!ht]
    \centering
    \includegraphics[width=1.0\linewidth]{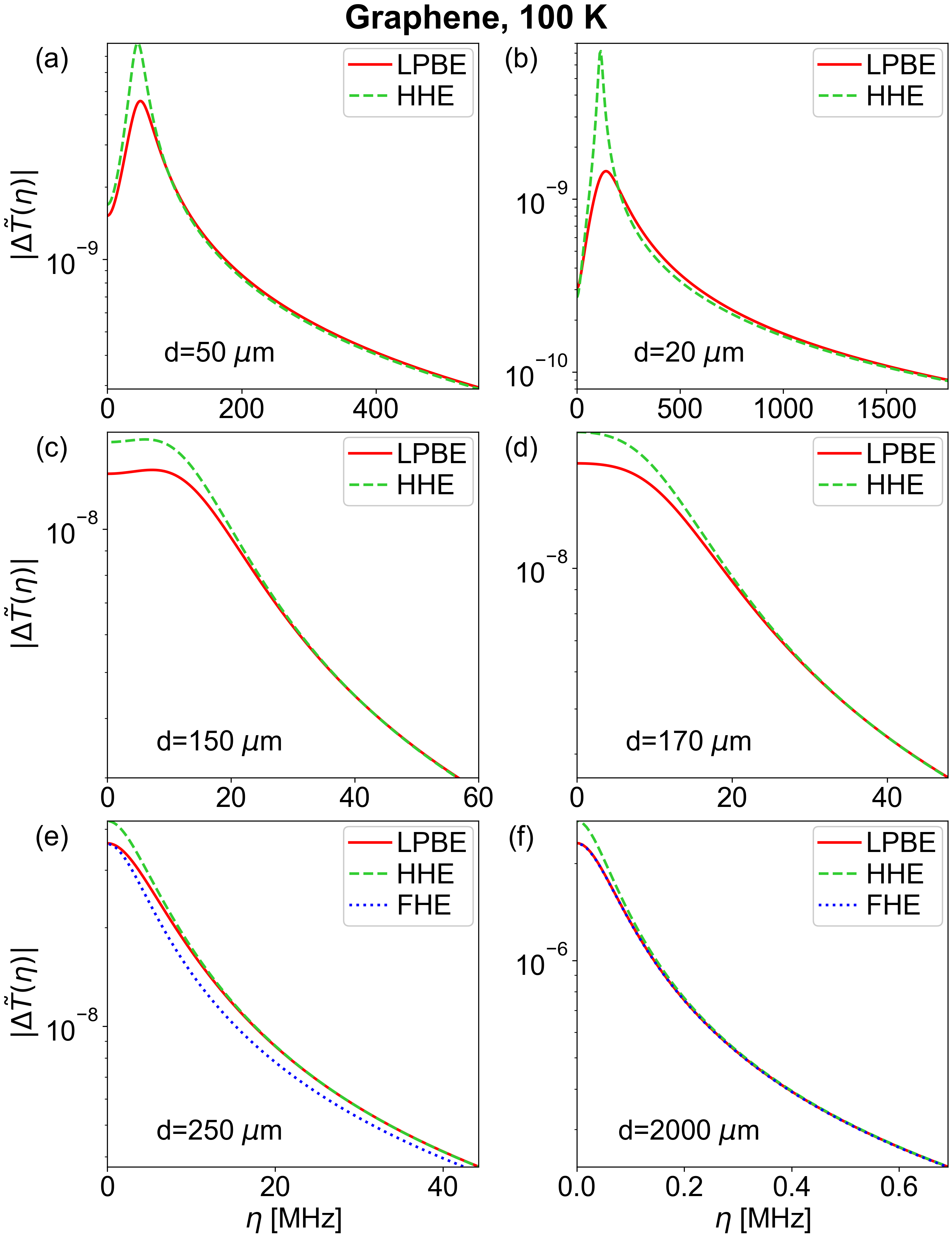}
    \caption{Absolute magnitudes of the temperature responses in the frequency domain [$| \Delta \tilde{T} \left( \eta > 0 \right)|$] corresponding to the time domain solutions shown in Fig.~\ref{fig:tg_soln_Graphene}, for grating periods ($d$) of (a) 50~$\mu$m, (b) 20~$\mu$m, (c) 150~$\mu$m, (d) 170~$\mu$m, (e) 250~$\mu$m and (f) 2000~$\mu$m. Here, we observe a peak at a non-zero $\eta$ for 50~$\mu$m, 20~$\mu$m and  150~$\mu$m grating periods, indicating an oscillatory temporal temperature response. The absence of this peak for the other grating periods rules out any oscillatory temporal responses. The non-oscillatory and non-exponential hydrodynamic temporal decay observed in Fig.~\ref{fig:tg_soln_Graphene} (e) exhibits a LPBE frequency domain response that matches with that of the HHE, but not with that of the FHE in Fig.~(e). For long grating periods [e.g., $d = 2000\ \mu$m shown in Fig.~(f)], the solutions of the LPBE and the FHE match exactly with each other, while the solution of HHE deviates from that of the LPBE closer to the steady-state ($\eta=0$) due to the 15\% lower effective steady-state $\kappa$ entering the former, as detailed in the main text.}
    \label{fig:freq_domain_tg_soln_Graphene}
\end{figure}

In Fig.~\ref{fig:freq_domain_tg_soln_Graphene}, we analyze the frequency domain responses corresponding to the time domain solutions shown in Fig.~\ref{fig:tg_soln_Graphene}. For the 50~$\mu$m grating period, both LPBE and HHE solutions show a narrow peak at non-zero frequency, indicating a single dominant oscillation frequency with a small bandwidth in the time domain response. This peak from the HHE becomes sharper at grating periods smaller than 50~$\mu$m, but wider at longer grating periods, as shown in Fig.~\ref{fig:freq_domain_tg_soln_Graphene}~(b) and~(c) for grating periods of 20~$\mu$m and 150~$\mu$m, respectively. However, for 170~$\mu$m grating period, this peak merges with the central peak ($\eta=0$), leading to vanishingly small oscillations in the time domain response. Further, for a 250~$\mu$m grating period, we observe a monotonically decreasing frequency response that deviates from the corresponding FHE solution, indicating a non-oscillatory, non-exponential temporal response. Lastly, for 2000~$\mu$m, the frequency response from the FHE overlaps with the LPBE response, confirming the Fourier-diffusive regime observed in Fig.~\ref{fig:tg_soln_Graphene}~(f).

It is important to note that the peak at $\eta \ne 0$ in the frequency domain solutions for the grating periods of 20~$\mu$m, 50~$\mu$m and  150~$\mu$m for graphene at 100 K shown in Fig.~\ref{fig:freq_domain_tg_soln_Graphene} are the global maximizers for $\left| \Delta \tilde{T} \left( \eta \right) \right|$, and therefore, are larger than the other local extrema observed for $\left| \Delta \tilde{T} \left( \eta \right) \right|$ at $\eta = 0$. We have shown in the Appendix~\ref{app_sec:frequency_analysis} that this feature is an essential requirement for the solution of the HHE to exhibit second sound-like oscillations. If, on the other hand, the peak at $\eta \ne 0$ is surpassed by the peak at $\eta = 0$, which occurs in the case of enriched boron nitride at 50 K (see Supplemental Material, section~S3), the resultant time domain temperature response ceases to show a clear negative dip and the oscillations are superimposed over a slowly decaying background, as shown in the Supplementary Fig. S29 (b).

\begin{figure}[!ht]
    \centering
    \includegraphics[width=0.85\linewidth]{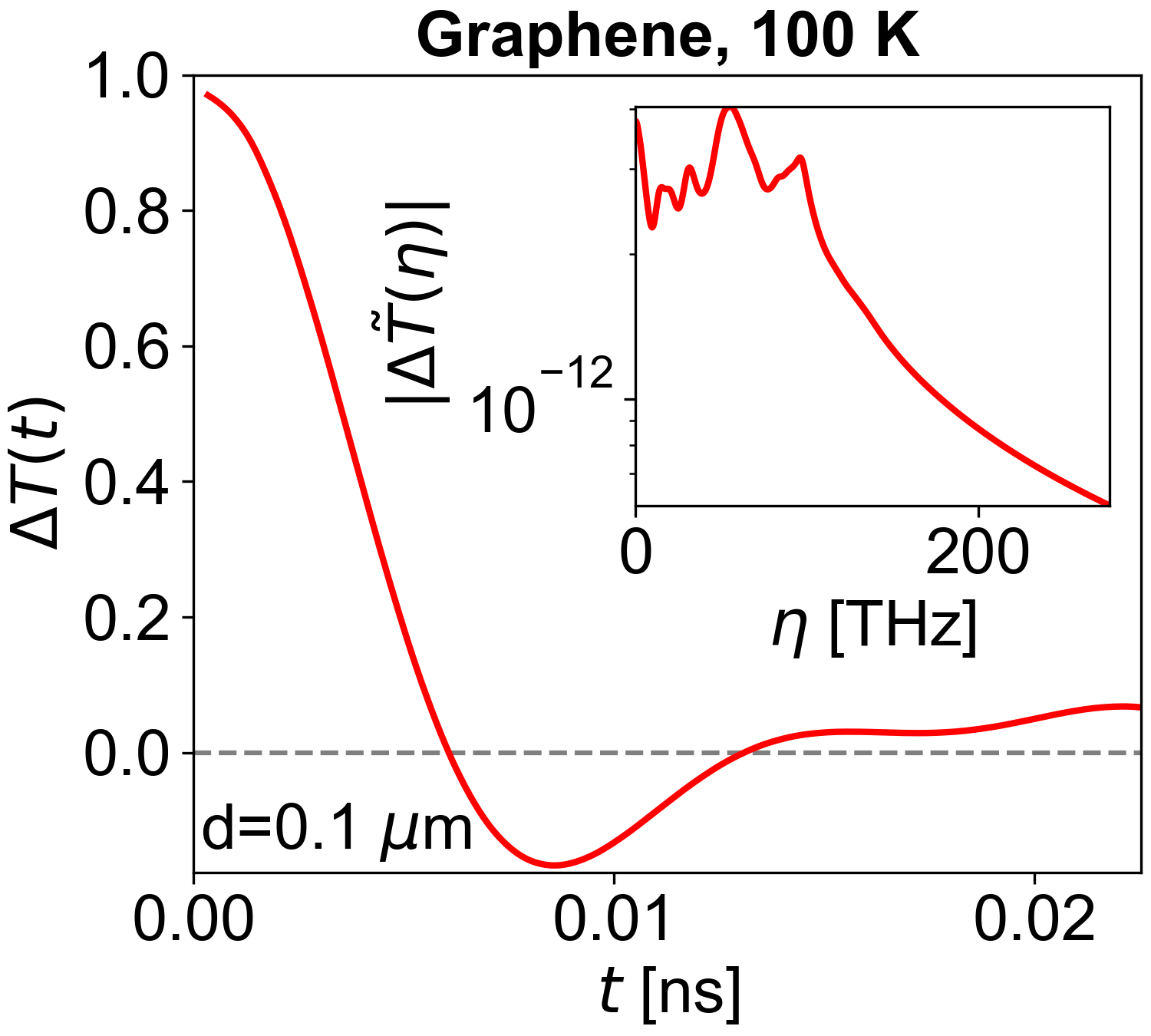}
    \caption{Transient temperature response for graphene at 100~K and a grating period of 0.1~$\mu$m. We observe an oscillatory temporal response, which corresponds to ballistic heat flow, as evidenced by the multiple peaks in the frequency domain response shown in the inset. Interestingly, the ballistic regime also exhibits a negative temperature dip similar to the hydrodynamic second sound regime described earlier, but the frequency domain responses show contrasting features between the two regimes, as discussed in the main text.}
    \label{fig:beyond_ddc_tg_soln_Graphene}
\end{figure}

This frequency domain analysis is pivotal for distinguishing oscillations due to hydrodynamic second sound from those due to ballistic heat flow. For the hydrodynamic second sound, the frequency domain solution exhibits a single peak at a non-zero frequency as discussed in the Appendix~\ref{app_sec:frequency_analysis} and shown in Fig.~\ref{fig:freq_domain_tg_soln_Graphene}~(a) for 50~$\mu$m grating period in graphene at 100~K. On the other hand, in the case of the ballistic heat flow regime, where all generalized Knudsen numbers exceed 1, the corresponding frequency domain solution exhibits multiple peaks as shown in Fig.~\ref{fig:beyond_ddc_tg_soln_Graphene} for graphene at 100~K and 0.1~$\mu$m grating period. Interestingly. the temporal temperature response exhibits a negative dip in the ballistic regime as well, as seen from Fig.~\ref{fig:beyond_ddc_tg_soln_Graphene}. Therefore, the observation of a single peak in the frequency response at a non-zero frequency along with a negative dip in the temporal temperature response, is a clear, unambiguous signature of hydrodynamic second sound in materials.

\section{Conclusions and discussion} \label{sec:discussion}
In summary, we demonstrate that a set of generalized spatial and temporal Knudsen numbers, originating from the spectral properties of the phonon collision matrix, determines the conditions for the Fourier-diffusive, hydrodynamic, weakly quasiballistic and ballistic heat flow regimes in semiconductors. These generalized Knudsen numbers enable a systematic reduction of the governing equation for phonon transport --- the linearized Peierls-Boltzmann equation (LPBE) --- into the continuum equations for the spatio-temporal evolution of the temperature field that characterize the different heat flow regimes. The limiting values of these generalized Knudsen numbers also inform the experimental conditions to observe the onset of transitions from one heat flow regime to another, without the need for the computationally intensive complete spatio-temporal solutions of the LPBE at different experimental conditions in a trial-and-error approach. We demonstrate the predictive capability of this generalized Knudsen number framework by confirming the predicted heat flow regimes from the continuum equations with the complete first-principles solution of the LPBE for twenty two different III-V and group IV semiconductors as well as for graphene under different experimental conditions. Our work provides a rigorous connect between the often-used continuum equations for different heat flow regimes and the fundamental microscopic governing equation for the transport of thermal phonons --- the LPBE --- in all non-magnetic semiconducting crystals, and elucidates important signatures of the dynamics of the continuum temperature field in each of these conventional Fourier-diffusive as well as the unconventional heat flow regimes beyond the Fourier's law, that will aid in their unambiguous experimental observations in the future.

In the context of the existing literature related to this work, several important findings have emerged out of our study, as we summarize below:
\begin{enumerate}
    \item \emph{Knudsen numbers from phonons vs. Knudsen numbers from the eigenmodes of $\bm{\Omega}$}: Spatial and temporal Knudsen numbers defined based on the total mean free path ($\Lambda_\lambda$) and total relaxation time ($\tau_\lambda$) of a phonon mode $\lambda$ as $\mathcal{K}^{\text{ph.}}_\lambda\left(\xi_x\right) \sim \xi_x\Lambda_\lambda$ and $\mathcal{K}^{\text{ph.}}_\lambda\left(\eta\right) \sim \eta\tau_\lambda$ respectively, where $\xi_x$ is the spatial wave vector and $\eta$ is the temporal frequency, have been used to identify the onset of weakly quasiballistic as well as ballistic heat flow regimes in materials like silicon~\cite{hua_transport_2014}, where the RTA describes heat flow reasonably well. Here, we have shown that the generalized Knudsen numbers derived from the properties of the eigenmodes of $\bm{\Omega}$ can predict the non-Fourier heat flow regimes, even for materials where the RTA is insufficient and a full LPBE description of heat flow is necessary. Furthermore, unlike the phonon Knudsen numbers, which are applicable for the non-hydrodynamic materials only, the generalized Knudsen numbers can be applied to hydrodynamic and non-hydrodynamic materials, and can predict the onset of hydrodynamic and ballistic heat flow regimes in the former as well as the weakly quasiballistic and ballistic heat flow regimes in the latter.
    \item \emph{Transient hydrodynamics is not always oscillatory}: Unsteady hydrodynamic heat flow is often considered synonymous with the oscillatory second sound regime~\cite{lee_hydrodynamic_2020}. While we do observe the oscillatory second sound regime in hydrodynamic materials like graphene at 100~K, we find that it occurs only within a narrow window of heating length scales. We have also identified, for the first time, a new non-oscillatory and non-exponential temporal decay of the temperature field that arises out of the transient hydrodynamic heat equation.
    \item \emph{Mutual exclusivity of the hydrodynamic and the weakly quasiballistic heat flow regimes}: As shown in Figs.~\ref{fig:regime_specific_condition_tg_soln_Si} (a) and~\ref{fig:regime_specific_condition_tg_soln_GaAs_InP} (a)-(b) as well as in the Supplemental Material, section S2, in all non-hydrodynamic materials, $\chi_{wQC} \ll \chi_{HGC}$. On the other hand, in hydrodynamic materials, $\chi_{wQC} \gtrsim \chi_{HGC}$, as we have shown for graphene at 100 K [Fig.~\ref{fig:regime_specific_condition_Graphene} in the main text], for BAs at 80 K [Supplementary Fig. S19 (a)], for enriched diamond at 100 K [Supplementary Fig. S27] and for enriched BN at 50-60 K [Supplementary Fig. S28 (a)-(b)]. Since $\chi_{HGC} \lesssim 1$ and $\chi_{wQC} \ll 1$ are the necessary conditions for the weakly quasiballistic heat flow regime, the regime classifiers for the hydrodynamic materials indicate that the weakly quasiballistic heat flow regime is unlikely to occur in them. We have already established that the hydrodynamic heat flow regime cannot occur in non-hydrodynamic materials. Thus, our findings lead to the interesting result of the mutual exclusivity of the two non-Fourier heat flow regimes in any semiconducting crystal.
    \item \emph{Unambiguous signatures of the hydrodynamic second sound regime}: We have shown in the Appendix~\ref{app_sec:frequency_analysis} that the oscillatory second sound regime must exhibit a single peak at a non-zero frequency ($\eta \ne 0$) in the frequency domain solution, and this peak must be a global maximizer. In particular, the other local extrema (at $\eta = 0$ and $\eta \to \infty$) must not be global maximizers. These specific features in the frequency domain solution result in an oscillatory time domain temperature response that dips below zero periodically, as shown for graphene at 100~K and a grating period of 50~$\mu$m in Fig.~\ref{fig:tg_soln_Graphene}~(a). However, the negative dip in the temporal temperature response occurs in the ballistic heat flow regime as well [see e.g., Fig.~\ref{fig:beyond_ddc_tg_soln_Graphene} for graphene at 100~K and a grating period of 0.1~$\mu$m]. Hence, the features of the frequency domain solution of the hydrodynamic second sound regime presented in Appendix~\ref{app_sec:frequency_analysis} are the unambiguous signatures of this unconventional heat flow regime in semiconductors.
    \item \emph{A characteristic heating length for the strongest hydrodynamic second sound signatures}: It has been shown in Ref.~\cite{hardy_phonon_1970} for cubic semiconductors that the hydrodynamic second sound oscillations cannot be observed when the spatial wave vectors are too large or too small. In this manuscript, we have shown using the regime classifiers as well as the full solutions of the LPBE that the findings of Ref.~\cite{hardy_phonon_1970} are more broadly applicable to two-dimensional materials as well after relaxing the requirements of the cubic symmetry. Furthermore, we have identified the region of intersection of the regime classifiers --- $\chi_{OC}$ and $\chi_{HGC}$ --- as the optimal spatial wave vectors (or equivalently, TG grating periods) to observe the strongest second sound signatures in a hydrodynamic material.
    \item \emph{Heating length-dependent second sound velocity within the hydrodynamic second sound regime}: In Ref.~\cite{ding_observation_2022}, a dependence of the second sound velocity ($v_{ss}$) with grating period was experimentally observed in graphite at 100~K to 150~K, and was attributed to the transition from hydrodynamic second sound to ballistic heat flow regime with decreasing grating period. The authors in Ref.~\cite{ding_observation_2022} showed that, at long grating periods, the observed $v_{ss}$ is lower than the undamped wave velocity ($v_{ud}$), which they refer to as the intrinsic second sound velocity, while at shorter grating periods, $v_{ss}$ exceeds $v_{ud}$ due to the onset of ballistic heat flow. In our work, we have shown that, in fact, the second sound velocity has a strong dependence on the grating wave vector even within the hydrodynamic second sound regime.
    \item \emph{Non-locality in heat flow}: Conventional Fourier's diffusion law is \emph{local} in nature, i.e., the heat flux at a point in space and time depends only on the temperature gradient at that same space-time point, i.e., $\bm{J}\left(\bm{x}, t\right) = -\kappa \nabla T\left(\bm{x}, t\right)$. Several works in the past have proposed a generalization of this \emph{local} Fourier's law to include non-local effects, effectively resulting in --- $\tilde{\bm{J}}\left(\bm{\xi}, \eta\right) = i\bm{\xi} \tilde{\kappa}\left(\bm{\xi}, \eta\right)\Delta\tilde{T}\left(\bm{\xi}, \eta\right)$ which gives a convolution in real space and time upon inverting the Fourier transform~\cite{hua_space-time_2020, chaput_direct_2013, kefayati_nonlocal_2022}. Starting from the expression for the heat flux, $\bm{J} = \left(1/V\right)\sum_{\lambda}\hbar\omega_\lambda \bm{v}_\lambda \sqrt{f_{\lambda}^{0} \left( f_{\lambda}^{0} + 1 \right)} f_\lambda' = \left(C_{0}/\zeta\right) \sum_{\alpha} \bm{\mathcal{V}}^{0 \alpha} \vartheta^{\alpha}$ and using the expression for $\bar{\vartheta}^\alpha$ from Eq.~\ref{eq:theta_alpha_Gen_sol}, we also arrive at a non-local expression for the generalized Fourier's law with $\tilde{\kappa}\left(\xi_x, \eta\right) = C_{0} \sum_{\alpha \beta} \mathcal{V}_{x}^{0 \alpha} \left[ \Gamma^{-1} \right]^{\alpha \beta} \mathcal{V}_{x}^{\beta 0}$. Interestingly, when $\mathcal{K}_{\alpha\beta} \ll 1$ for all $\alpha\ne\beta$ and $\mathcal{K}_t \ll 1$, we obtain $\tilde{\kappa}\left(\xi_x, \eta\right) \approx \tilde{\kappa}\left(\xi_x\right) = \sum_\alpha \mathcal{S}_\alpha \kappa_\alpha$ and a spatially non-local generalized Fourier's law --- consistent with the weakly quasiballistic regime --- emerges. On the other hand, when $\mathcal{K}_{\alpha\beta} \ll 1$ for all $\left(\alpha, \beta\right)$ and the HTC requirement on $\mathcal{K}_t$ is satisfied --- the requirements for hydrodynamic heat flow in a hydrodynamic material, a temporally non-local generalized Fourier's law emerges. Thus, our work also bridges the non-local generalizations of the Fourier's law with the continuum descriptions of the non-Fourier heat flow regimes --- particularly the wQHE and the HHE.
    \item \emph{Controlling spatial wave vectors vs. temporal frequencies}: Among the two parameters in the LPBE $\left(\xi, \eta\right)$ that determine the nature of the solution, experiments typically have explicit control over at most one of them (e.g., by choosing the grating period for the heat source in TG) while the other manifests in the response to the imposed heat source, and is often interpreted as an appropriate thermal scale (e.g., the thermal frequency scale, $\eta \sim \rho_sk^2$). In our work, we have chosen the TG experiment to elucidate our findings, since the weakly quasiballistic~\cite{johnson_direct_2013, ravichandran_spectrally_2018} and the hydrodynamic second sound~\cite{huberman_observation_2019, ding_observation_2022, xie_room-temperature_2026} regimes have been observed with this experiment in the recent past. Therefore, the generalized spatial Knudsen numbers are controlled externally for a given material and the generalized temporal Knudsen numbers are inferred from the knowledge of the thermal frequency scale. It is trivial to extend our analysis to experiments where the generalized temporal Knudsen numbers are externally controlled and the generalized spatial Knudsen numbers are inferred from the knowledge of the thermal length scale, e.g., in situations where the surface of a crystal is heated in a temporally sinusoidal but spatially uniform manner. In fact, such experiments offer interesting pathways to investigate the HTC systematically by varying $\eta$ in a hydrodynamic material, which is not possible in a TG set-up.  
\end{enumerate}

\appendix
\section{First principles solution of the linearized Peierls-Boltzmann equation for phonon transport} \label{app_sec:first_principles_desciprion}
For this work, we calculate the harmonic and the anharmonic phonon properties, and the phonon collision matrix ($\bm{\Omega}$) as discussed in Refs.~\cite{ravichandran_unified_2018, ravichandran_phonon-phonon_2020, malviya_failure_2023} and briefly summarized in the Supplemental Material, section~S3. Further, for graphene, we employ the anharmonic renormalization of the ZA phonon dispersions that is necessary to stabilize the flat phase of suspended two-dimensional materials as discussed in Refs.~\cite{ravichandran_low_2026, ravichandran_elasticity_2026}. For the calculation of $\bm{\Omega}$, we have considered three-phonon scattering processes for isotopically pure samples of the III-V and group IV compounds (and also included phonon-isotope scattering for naturally occurring samples), since the effect of higher-order scattering among four phonons on the $\kappa$ of these materials is very weak at the temperatures considered here~\cite{ravichandran_phonon-phonon_2020}. However, for graphene, four-phonon scattering has been found to strongly affect the $\kappa$ even at temperatures around and below 150~K in recent works~\cite{ravichandran_elasticity_2026, tur-prats_high-order_2025, li_effects_2025}, and so, has been included in our calculations. Further, the calculated eigenmodes of $\bm{\Omega}$ do not exhibit exact even-odd parity due to the finite numerical precision of the computations; therefore, to classify these eigenmodes, we find their contribution to $\kappa$. If they are small ($\kappa_{\alpha} < 10^{-4} \times \max_{\alpha} (\kappa_{\alpha})/\kappa$), we classify them as an even; otherwise, they are odd. Finally, we use the low-rank method discussed in Ref.~\cite{malviya_efficient_2025} to obtain the transient solution of LPBE. Here, we used nearly 36\%, 54\%, and 70\% of eigenmodes as required to get 99\% of the total $\kappa$, for the LPBE solution in the cases of Si, GaAs, and InP, respectively at 100~K. Whereas in the case of graphene at 100~K, we require only $\sim$9\% of the eigenmodes of $\bm{\Omega}$ to obtain $\sim$99.7\% of total $\kappa$. For all of these cases, the numerical discretization density of the Brillouin zone that is necessary to obtain the converged properties of the eigenmodes of $\bm{\Omega}$~\cite{malviya_indicators_2026} are tabulated in the Supplemental Material, section S3. \\

\section{Conditions on generalized Knudsen numbers for different heat flow regimes} \label{app_sec:genralized_regime_classifiers}
Here, we derive the necessary conditions on the generalized Knudsen numbers (introduced in the main text) to reduce the GHE [Eq.~\ref{eq:lpbe_frequency_domain_temperature_response}] to the IHE [Eq.~\ref{eq:intermediate_frequency_domain_temperature_response}], then subsequently to the FHE [Eq.~\ref{eq:fourier_diffusion_heat_equation}], the wQHE [Eq.~\ref{eq:quasiballistic_heat_equation}] and the HHE [Eq.~\ref{eq:hyperbolic_heat_equation}].

\subsection{Diagonal dominance condition (DDC)} \label{app_subsec:diagonal_dominance_condition}
As discussed in section~\ref{sec:generalized_regime_classifiers}, for the diagonal dominance of $\mathbf{\Gamma}$ we require $\max_{\beta,\gamma\ne\beta}\left|\Gamma_{od}^{\beta \gamma}\right| \sqrt{\left|\left[ \Gamma_{d}^{-1} \right]^{\beta \beta}\right|\left|\left[ \Gamma_{d}^{-1} \right]^{\gamma \gamma}\right|} \ll 1$, where: 
\begin{subequations}
\begin{align}
    \Gamma^{\beta \gamma}_{d} & = \left[ -i \eta + \sigma^{\beta} + \sum_{\bar{\alpha}>0} \frac{ \left( \xi_{x} \mathcal{V}^{\beta \bar{\alpha}}_{x} \right)^{2}}{\sigma^{\bar{\alpha}}-i \eta} \right] \Delta_{\beta \gamma}
    \label{eq:Gamma_diagonal} \\
    \Gamma^{\beta \gamma}_{od} & = \sum_{\bar{\alpha}>0} \frac{ \left( \xi_{x} \mathcal{V}^{\beta \bar{\alpha}}_{x} \right) \left( \xi_{x} \mathcal{V}^{\bar{\alpha} \left( \gamma \neq \beta \right) }_{x} \right) }{\sigma^{\bar{\alpha}}-i \eta}.
    \label{eq:Gamma_non_diagonal}
\end{align}
\end{subequations}

To arrive at the properties of the eigenmodes of $\bm{\Omega}$ and the Fourier variables $\left(\eta, \xi_x\right)$ that satisfy the above requirements, we note that, 
\begin{align*}
    \left| \Gamma^{\gamma \gamma}_{d} \right| > & \mathcal{R} \left( \Gamma^{\gamma \gamma}_{d} \right) = \sigma^{\gamma} + \sum_{\bar{\alpha}>0} \frac{ \sigma^{\bar{\alpha}} \left( \xi_{x} \mathcal{V}^{ \gamma \bar{\alpha}}_{x} \right)^{2} }{\left(\sigma^{\bar{\alpha}}\right)^{2} + \eta^{2}} > \sigma^{\gamma}  \\
    \implies \left| \Gamma^{-1}_{d} \right|^{\gamma \gamma} < & \frac{1}{\sigma^{\gamma}}. 
\end{align*}
where, $\mathcal{R} \left( \Gamma^{\gamma \gamma}_{d} \right)$ is the real part of $\Gamma^{\gamma \gamma}_{d}$. Furthermore,
\begin{align*}
    \left| \Gamma^{\beta \left(\gamma\ne\beta\right)}_{od} \right| & = \left| \sum_{\bar{\alpha}>0} \frac{\left( \xi_{x} \mathcal{V}^{\beta \bar{\alpha}}_{x} \right) \left( \xi_{x} \mathcal{V}^{\bar{\alpha} \left( \gamma \neq \beta \right)}_{x} \right) }{\sigma^{\bar{\alpha}} -i \eta } \right| \nonumber \\
    & \leq \sum_{\bar{\alpha}>0} \frac{\left| \left( \xi_{x} \mathcal{V}^{\beta \bar{\alpha}}_{x} \right) \left( \xi_{x} \mathcal{V}^{\bar{\alpha} \left( \gamma \neq \beta \right)}_{x} \right) \right|}{\sqrt{\left(\sigma^{\bar{\alpha}}\right)^{2} + \eta^{2}}} \nonumber \\
    & < \sum_{\bar{\alpha}>0} \frac{\left| \left( \xi_{x} \mathcal{V}^{\beta \bar{\alpha}}_{x} \right) \left( \xi_{x} \mathcal{V}^{\bar{\alpha} \left( \gamma \neq \beta \right)}_{x} \right) \right|}{\sigma^{\bar{\alpha}}}
\end{align*}
where we have used the triangle inequality in the second step. Therefore, the requirement $\left|\Gamma_{od}^{\beta \gamma}\right| \sqrt{\left|\left[ \Gamma_{d}^{-1} \right]^{\beta \beta}\right|\left|\left[ \Gamma_{d}^{-1} \right]^{\gamma \gamma}\right|} \ll 1$ for all $\beta, \gamma\ne\beta > 0$ becomes:
\begin{align}
    \chi_{DDC} = \max_{\beta, \gamma \ne \beta > 0}\mathcal{K}_{\beta\gamma}^2 \ll 1
\end{align}
where $\mathcal{K}_{\beta\gamma}^2 = \sum_{\bar{\alpha}>0} \frac{\left| \left( \xi_{x} \mathcal{V}^{\beta \bar{\alpha}}_{x} \right) \left( \xi_{x} \mathcal{V}^{\bar{\alpha} \left( \gamma \neq \beta \right)}_{x} \right) \right|}{\displaystyle \sqrt{\sigma^{\beta}\sigma^{\bar{\alpha}}} \sqrt{\sigma^{\beta}\sigma^{\gamma}} }$

\subsection{Weakly quasiballistic condition (wQC)} \label{app_subsec:weakly_quasiballistic_condition_proof}
To obtain the wQC, we start from the intermediate heat equation (IHE, Eq.~\ref{eq:intermediate_frequency_domain_temperature_response}), which is rewritten by splitting $\mathbf{\Gamma}_{d}$ into its real and imaginary parts as:
\begin{align}
    -i \left( \eta + \sum_{\alpha > 0} \frac{\left( \xi_{x} \mathcal{V}_{x}^{0 \alpha} \right)^{2} \mathcal{I}}{\mathcal{I}^{2} + \mathcal{R}^{2}} \right) \Delta \tilde{T} & \nonumber \\
    + \sum_{\alpha > 0} \frac{\left( \xi_{x} \mathcal{V}^{ 0 \alpha}_{x} \right)^{2} \mathcal{R}}{\mathcal{I}^{2} + \mathcal{R}^{2}} \Delta \tilde{T} & = \frac{\tilde{h}^{0}}{\zeta}
    \label{eq:intermediate_frequency_domain_temperature_response_1}
\end{align}
where, $\mathcal{R}$ and $\mathcal{I}$ are the real and imaginary parts of $\Gamma_{d}^{\alpha \alpha}$ respectively, given by:
\begin{align*}
    \mathcal{R} & = \sigma^{\alpha} + \sum_{\bar{\beta} > 0} \frac{\left( \xi_{x} \mathcal{V}_{x}^{\alpha \bar{\beta}} \right)^{2} \sigma^{\bar{\beta}}}{\left( \sigma^{\bar{\beta}} \right)^{2} + \eta^{2}} \\
    \mathcal{I} & = - \eta \left[ 1 + \sum_{\bar{\beta} > 0} \frac{\left( \xi_{x} \mathcal{V}_{x}^{\alpha \bar{\beta}} \right)^{2}}{\left( \sigma^{\bar{\beta}} \right)^{2} + \eta^{2}} \right].
\end{align*}

When $\chi_{wQC}\left(\eta\right) = \mathcal{K}_t \ll 1$, where $\mathcal{K}_t = \frac{\eta_T}{\sigma_{\text{min.}}}$ with $\sigma_{\text{min.}} = \min \left( \sigma \ne \sigma^0 \right)$ and $\eta_T = \sum_{\alpha}\frac{\xi_x^2\left(\mathcal{V}^{0\alpha}\right)^2}{\sigma^\alpha}$ being the thermal decay frequency taken, conservatively, to be that corresponding to a diffusive heat flow regime, we can approximate $( \sigma^{\bar{\beta}} )^{2} + \eta^{2} \approx ( \sigma^{\bar{\beta}} )^{2}$, since all relevant temporal frequencies ($\eta$) in the time domain solution, $\Delta T\left(x, t\right)$, will be comparable to or smaller than $\eta_T$, even when the actual heat flow regime in non-diffusive in nature~\cite{hua_transport_2014}. Thus, $\mathcal{R}$ and $\mathcal{I}$ can be simplified as:
\begin{align*}
    \mathcal{R} & \approx \sigma^{\alpha} + \sum_{\bar{\beta} > 0} \frac{\left( \xi_{x} \mathcal{V}_{x}^{\alpha \bar{\beta}} \right)^{2}}{\sigma^{\bar{\beta}}} \text{, and} \\
    \mathcal{I} & \approx - \eta \left[ 1 + \sum_{\bar{\beta} > 0} \frac{\left( \xi_{x} \mathcal{V}_{x}^{\alpha \bar{\beta}} \right)^{2}}{\left( \sigma^{\bar{\beta}} \right)^{2}} \right]
\end{align*}

Furthermore, $ -\mathcal{I} \ll \mathcal{R}$ since the individual terms of $-\mathcal{I}$ are significantly smaller than those of $\mathcal{R}$. Therefore, in Eq.~\ref{eq:intermediate_frequency_domain_temperature_response_1}, we approximate $\mathcal{I}^{2} + \mathcal{R}^{2} \approx \mathcal{R}^{2}$ to get:
\begin{align}
    -i \eta \left( 1 + \frac{1}{\eta} \sum_{\alpha > 0} \frac{\left( \xi_{x} \mathcal{V}_{x}^{0 \alpha} \right)^{2} \mathcal{I}}{\mathcal{R}^{2}} \right) \Delta \tilde{T} & \nonumber \\
    + \sum_{\alpha > 0} \frac{\left( \xi_{x} \mathcal{V}^{ 0 \alpha}_{x} \right)^{2}}{\mathcal{R}} \Delta \tilde{T} & = \frac{\tilde{h}^{0}}{\zeta}
    \label{eq:intermediate_frequency_domain_temperature_response_2}
\end{align}
This equation takes the form of the weakly quasiballistic heat equation (wQHE) [Eq.~\ref{eq:fourier_like_frequency_domain_temperature_response}] when:
\begin{align}
    & \chi_{R-wQC}  \left( \xi_{x} \right) = \frac{1}{\eta} \sum_{\alpha > 0} \frac{\left( \xi_{x} \mathcal{V}_{x}^{0 \alpha} \right)^{2} \left| \mathcal{I} \right| }{ \mathcal{R}^{2} } \nonumber \\
    & = \xi_{x}^{2} \sum_{\alpha > 0} \left[ \frac{\mathcal{V}^{0 \alpha}}{\sigma^{\alpha}} \right]^{2} \frac{\left[ 1 + \sum_{\bar{\beta} > 0} \frac{\left( \xi_{x} \mathcal{V}_{x}^{\alpha \bar{\beta}} \right)^{2}}{\left( \sigma^{\bar{\beta}} \right)^{2}} \right]}{\left[ 1 + \sum_{\bar{\beta} > 0} \frac{\left( \xi_{x} \mathcal{V}_{x}^{\alpha \bar{\beta}} \right)^{2}}{\sigma^{\alpha} \sigma^{\bar{\beta}}} \right]^{2}} \nonumber\\
    &= \sum_{\alpha > 0}\frac{ \frac{\xi_{x}^{2} \left[\mathcal{V}^{0 \alpha}\right]^2}{\sigma^{\alpha}} }{\sigma^{\alpha}} \frac{\left[ 1 + \sum_{\bar{\beta} > 0} \frac{\left( \xi_{x} \mathcal{V}_{x}^{\alpha \bar{\beta}} \right)^{2}}{\left( \sigma^{\bar{\beta}} \right)^{2}} \right]}{\left[ 1 + \sum_{\bar{\beta} > 0} \frac{\left( \xi_{x} \mathcal{V}_{x}^{\alpha \bar{\beta}} \right)^{2}}{\sigma^{\alpha} \sigma^{\bar{\beta}}} \right]^{2}} \nonumber\\
    &\leq \sum_{\alpha > 0}\frac{ \frac{\xi_{x}^{2} \left[\mathcal{V}^{0 \alpha}\right]^2}{\sigma^{\alpha}} }{\sigma^{\alpha}} \frac{\left[ 1 + \sum_{\bar{\beta} > 0} \frac{\left( \xi_{x} \mathcal{V}_{x}^{\alpha \bar{\beta}} \right)^{2}}{\left( \sigma^{\bar{\beta}} \right)^{2}} \right]}{\left[ 1 + \sum_{\bar{\beta} > 0} \frac{\left( \xi_{x} \mathcal{V}_{x}^{\alpha \bar{\beta}} \right)^{2}}{\sigma^{\alpha} \sigma^{\bar{\beta}}} \right]} \nonumber\\
    &\leq \sum_{\alpha > 0}\frac{ \frac{\xi_{x}^{2} \left(\mathcal{V}^{0 \alpha}\right)^2}{\sigma^{\alpha}} }{\sigma_{\text{min.}}} \frac{\left[ \frac{\sigma_{\text{min.}}}{\sigma^\alpha} + \sum_{\bar{\beta} > 0} \frac{\sigma_{\text{min.}}}{\sigma^{\bar{\beta}}}\frac{\left( \xi_{x} \mathcal{V}_{x}^{\alpha \bar{\beta}} \right)^{2}}{\sigma^\alpha \sigma^{\bar{\beta}} } \right]}{\left[ 1 + \sum_{\bar{\beta} > 0} \frac{\left( \xi_{x} \mathcal{V}_{x}^{\alpha \bar{\beta}} \right)^{2}}{\sigma^{\alpha} \sigma^{\bar{\beta}}} \right]} \ll 1
    \label{eq:fourier_like_condition_2}
\end{align}

Each term within the square bracket in the numerator in Eq.~\ref{eq:fourier_like_condition_2} is less that the corresponding term in the denominator, since $\sigma_{\text{min.}}/\sigma^\alpha \leq 1$ and $\sigma_{\text{min.}}/\sigma^{\bar{\beta}} \leq 1$ for all $\alpha$ and $\bar{\beta}$. Therefore, $\chi_{R-wQC} < \sum_{\alpha > 0} \frac{ \frac{\xi_{x}^{2} \left(\mathcal{V}^{0 \alpha}\right)^2}{\sigma^{\alpha}} }{\sigma_{\text{min.}}} = \chi_{wQC}$. Therefore, the requirement of $\chi_{R-wQC} \ll 1$ is always satisfied when the weakly quasiballistic condition, $\chi_{wQC} \ll 1$ is satisfied. Hence, we refer to this condition, $\chi_{R-wQC} \ll 1$, as the redundant weakly quasiballistic condition (R-wQC). 

Thus, when $\chi_{wQC} \ll 1$, Eq.~\ref{eq:intermediate_frequency_domain_temperature_response_2} takes the form of the following diffusion equation:
\begin{align*}
    -i \eta \Delta \tilde{T} + \xi_{x} ^{2} \left[ \sum_{\alpha > 0} \frac{\left( \mathcal{V}^{ 0 \alpha}_{x} \right)^{2}}{\mathcal{R}} \right] \Delta \tilde{T} & = \frac{\tilde{h}^{0}}{\zeta}
\end{align*}
Here, the term within square brackets is the suppressed thermal diffusivity ($\rho_{s}$), which expands as:
\begin{align*}
    \rho_{s} = \sum_{\alpha > 0} \frac{\left( \mathcal{V}^{ 0 \alpha}_{x} \right)^{2}}{\mathcal{R}} = \sum_{\alpha > 0} \frac{\left( \mathcal{V}^{ 0 \alpha}_{x} \right)^{2}}{\sigma^{\alpha}} \underbrace{\left[ 1 + \sum_{\bar{\beta} > 0} \frac{\left( \xi_{x} \mathcal{V}_{x}^{\alpha \bar{\beta}} \right)^{2}}{\sigma^{\alpha} \sigma^{\bar{\beta}}} \right]^{-1}}_{\mathcal{S}_\alpha}.
\end{align*} 
where $\mathcal{S}_\alpha$ is the weakly quasiballistic suppression function defined earlier.

We note that we would have arrived at the same conclusions (i.e., $\chi_{R-wQC}\left(\xi_x\right) \ll 1$ whenever $\chi_{wQC}\left(\eta\right) \ll 1$) from Eq.~\ref{eq:fourier_like_condition_2} even if we had chosen the thermal decay frequency, $\eta_T$, corresponding to the suppressed diffusivity while introducing the requirement of $\chi_{wQC}\left(\eta\right) \ll 1$, i.e., $\eta^s_T = \sum_\alpha \mathcal{S}_\alpha \frac{\xi_x^2\left(\mathcal{V}^{0\alpha}_x\right)^2}{\sigma^\alpha}$. Since $\eta^s_T$ is a more realistic thermal decay frequency in the non-Fourier heat flow regime and reduces to $\eta_T$ in the Fourier-diffusive heat flow regime, we use $\eta^s_T$ for checking the wQC requirement, $\chi_{wQC}\left(\eta\right) \ll 1$, in the main text.

\subsection{Hydrodynamic geometry and material conditions (HGC and HMC)} \label{app_subsec:hydrodynamic_second_sound_condition_proof}
As discussed in Ref~\cite{hardy_phonon_1970}, to reduce the IHE [Eq.~\ref{eq:intermediate_frequency_domain_temperature_response}] to the HHE [Eq.~\ref{eq:hyperbolic_heat_equation}], the diagonal part of the $\bm{\Gamma}$ matrix must be reduced as: $\Gamma_{d}^{\alpha \alpha} \approx  -i \eta + \sigma^{\alpha}$. To obtain this reduction, we consider the real ($\mathcal{R}$) and imaginary ($\mathcal{I}$) components of $\Gamma_{d}^{\alpha \alpha}$, as defined in Section~\ref{app_subsec:weakly_quasiballistic_condition_proof}, which follow the following inequalities:
\begin{align*}
    \mathcal{R} = \sigma^{\alpha} + \sum_{\bar{\beta}>0} \sigma^{\bar{\beta}} \frac{ \left( \xi_{x} \mathcal{V}^{ \alpha \bar{\beta}}_{x} \right)^{2}}{\left(\sigma^{\bar{\beta}}\right)^{2} + \eta^{2}} < \sigma^{\alpha} + \sum_{\bar{\beta}>0} \frac{ \left( \xi_{x} \mathcal{V}^{ \alpha \bar{\beta}}_{x} \right)^{2}}{\sigma^{\bar{\beta}}}
\end{align*}
and
\begin{align*}
    - \mathcal{I} = \eta + \eta \sum_{\bar{\beta}>0} \frac{ \left( \xi_{x} \mathcal{V}^{ \alpha \bar{\beta}}_{x} \right)^{2}}{\left(\sigma^{\bar{\beta}}\right)^{2} + \eta^{2}} < \eta +  \eta \sum_{\bar{\beta}>0} \frac{ \left( \xi_{x} \mathcal{V}^{ \alpha \bar{\beta}}_{x} \right)^{2}}{\left(\sigma^{\bar{\beta}}\right)^{2}}
\end{align*}

Under the following conditions:
\begin{subequations}
\begin{gather}
    \label{eq:hgc}
    \sum_{\bar{\beta}>0} \frac{ \left( \xi_{x} \mathcal{V}^{ \alpha \bar{\beta}}_{x} \right)^{2}}{\sigma^{\bar{\beta}} \sigma^{\alpha}} \ll 1 \\
    \quad \text{and } \quad \sigma^{\alpha} < \sigma^{\bar{\beta}}
\end{gather}    
\end{subequations}
We have:
\begin{align*}
    \sum_{\bar{\beta}>0} \frac{ \left( \xi_{x} \mathcal{V}^{ \alpha \bar{\beta}}_{x} \right)^{2}}{\left(\sigma^{\bar{\beta}}\right)^{2}} < \sum_{\bar{\beta}>0} \frac{ \left( \xi_{x} \mathcal{V}^{ \alpha \bar{\beta}}_{x} \right)^{2}}{\sigma^{\bar{\beta}} \sigma^{\alpha}} \ll 1
\end{align*}
which simplifies $\mathcal{R}$ and $\mathcal{I}$ as:
\begin{align*}
    \mathcal{R} \approx \sigma^{\alpha} \text{, and } \mathcal{I} \approx - \eta
\end{align*}
thus reducing the diagonal part of the $\bm{\Gamma}$ matrix as: $\Gamma_{d}^{\alpha \alpha} \approx -i \eta + \sigma^{\alpha}$, which reduces the IHE [Eq.~\ref{eq:intermediate_frequency_domain_temperature_response}] to:
\begin{align}
    - i \eta \Delta \tilde{T} + \sum_{\alpha > 0} \frac{\left( \xi_{x} \mathcal{V}_{x}^{0 \alpha} \right)^{2}}{- i \eta + \sigma^{\alpha}} \Delta \tilde{T} = \frac{\tilde{h}}{\zeta}
    \label{eq:intermediate_hydrodynamics_second_sound}
\end{align}
Next, we introduce the additional requirement that there must exist a group of eigenmodes belonging to a set $\mathcal{D}$ such that:
\begin{align}
    \sum_{\alpha \in \mathcal{D}} \frac{\left( \xi_{x} \mathcal{V}^{ 0 \alpha}_{x} \right)^{2}}{\left| -i \eta + \sigma^{\alpha}\right|} \gg \sum_{\alpha \notin \mathcal{D}} \frac{\left( \xi_{x} \mathcal{V}^{ 0 \alpha}_{x} \right)^{2}}{\left| -i \eta + \sigma^{\alpha}\right|}.
    \label{eq:dominating_eigenmode_condition}
\end{align}
and the requirement that these eigenmodes must have nearly degenerate eigenvalues ($\sigma^{i} \approx \mu$, for all $i \in \mathcal{D}$), which further reduces Eq.~\ref{eq:dominating_eigenmode_condition} to:
\begin{align}
    \sum_{\alpha \in \mathcal{D}} \frac{\left( \xi_{x} \mathcal{V}^{ 0 \alpha}_{x} \right)^{2}}{\sqrt{\eta^2 + \mu^2}} &\gg \sum_{\alpha \notin \mathcal{D}} \frac{\left( \xi_{x} \mathcal{V}^{ 0 \alpha}_{x} \right)^{2}}{\sqrt{\eta^2 + \left(\sigma^{\alpha}\right)^2}}\nonumber\\
    \implies \frac{\sum_{\alpha\in\mathcal{D}}\kappa_\alpha}{\sqrt{\mathcal{K}_t^2 + 1}} &\gg \sum_{\alpha\in\mathcal{D}}\frac{\kappa_\alpha}{\sqrt{\mathcal{K}_t^2\left(\mu/\sigma^\alpha\right)^2 + 1}}
    \label{eq:reduced_dominating_eigenmode_condition}
\end{align}
where $\mathcal{K}_t = \eta/\mu$ as defined in the Appendix section~\ref{app_subsec:weakly_quasiballistic_condition_proof}. These additional requirements result in the following simplification for the second (summation) term in Eq.~\ref{eq:intermediate_hydrodynamics_second_sound}:
\begin{align*}
    \sum_{\alpha > 0} \frac{\left( \xi_{x} \mathcal{V}_{x}^{0 \alpha} \right)^{2}}{- i \eta + \sigma^{\alpha}} \approx \xi_{x}^{2}\frac{\sum_{\alpha \in \mathcal{D}} \left( \mathcal{V}^{ 0 \alpha}_{x} \right)^{2}}{-i \eta + \mu}
\end{align*}
which simplifies Eq.~\ref{eq:intermediate_hydrodynamics_second_sound} into the HHE in the frequency domain as given in Eq.~\ref{eq:second_sound_frequency_domain_temperature_response}. We note that Eq.~\ref{eq:reduced_dominating_eigenmode_condition}, in the steady-state limit of $\eta = 0$, reduces to:
\begin{align*}
    \sum_{\alpha \in \mathcal{D}} \kappa_{\alpha} \gg \sum_{\alpha \notin D} \kappa_{\alpha}.
\end{align*}
which is a strictly material condition that is necessary for the reduction of the IHE to the HHE. Here, $\kappa_\alpha = C_o\frac{\left(\mathcal{V}^{0\alpha}\right)^2}{\sigma^\alpha}$ is the thermal conductivity of the eigenmode $\alpha$. Thus, the first step in predicting the possibility of hydrodynamic heat flow in a material is to check for dominating contributions to $\kappa$ from eigenmodes with nearly-degenerate eigenvalues. This requirement is consistent with the limiting behavior of $\kappa$ when momentum-dissipative U-processes are vanishingly small (i.e., $\bm{\Omega} \approx \bm{\Omega}_N$), where three (two) eigenmodes of $\bm{\Omega}$ contribute entirely to $\kappa$ in three-(two-) dimensional materials~\cite{pitaevskii_physical_2012}.

\section{Frequency domain analysis of the continuum heat equation} \label{app_sec:frequency_analysis}
\begin{figure}[!ht]
    \centering
    \includegraphics[width=1.0\linewidth]{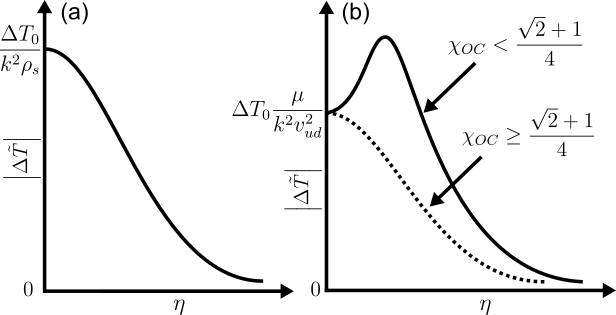}
    \caption{A schematic illustrating the absolute value of the frequency domain temperature response for TG setup. Here we have only shown the positive $\eta$ region due to its symmetry about $\eta=0$. (a) For the quasiballistic heat flow, the frequency domain solution is a monotonically decreasing bell-shaped function, centered at $\eta=0$. (b) For the hydrodynamic second sound, the frequency domain solution exhibits one side peak when $\chi_{oc} < (\sqrt{2}+1)/4$, while for $\chi_{oc} \geq (\sqrt{2}+1)/4$ representing the regimes of hydrodynamic heat flow with oscillations being incomplete or completely absent as described in the main text, the frequency domain response is a monotonically decreasing function.}
    \label{fig:schematic_fourier_analysis}
\end{figure}

In this section, we discuss the distinguishing features of the hydrodynamic, the weakly quasiballistic and the ballistic heat flow regimes by analyzing the absolute values of the temperature response in the frequency domain for a fixed spatial wave vector $\xi_x = 2\pi/d$ (corresponding to a TG grating period, $d$). In the weakly quasiballistic regime, the corresponding solution to the Fourier transform of the wQHE [Eq.~\ref{eq:fourier_like_frequency_domain_temperature_response}] at a fixed $\xi_x$ is given by: 

\begin{align}
    \left| \Delta \tilde{T} \left( \eta \right) \right| = \frac{\Delta T_{0}}{ \sqrt{ \left( k^{2} \rho_{s}\right)^{2} + \eta^{2}} }
\end{align}
As shown in Fig.~\ref{fig:schematic_fourier_analysis}~(a), this function has a bell-shaped form with zero slope at $\eta = 0$ and $\eta = \pm \infty$, and the ordinate at $\eta=0$ gives the thermal decay timescale $\left[1/\left(k^2\rho_s\right)\right]$.

For the hydrodynamic heat flow, the Fourier transformed temperature deviation ($|\Delta \tilde{T} |$)  from Eq.~\ref{eq:second_sound_TG_frequency_domain_temperature_response} is given by:
\begin{align}
    \left| \Delta \tilde{T} \left( \eta \right) \right| = \Delta T_{0} \sqrt{ \frac{\eta^{2} + \mu^{2}}{ \left( k^{2} v_{ud}^{2} - \eta^{2} \right)^{2} + \mu^{2} \eta^{2}} }
\end{align}
As shown in Fig.~\ref{fig:schematic_fourier_analysis}~(b), this function also has a vanishing slope at $\eta_{0} = 0$ and $\eta_{\infty} = \pm \infty$. Furthermore, the derivative of $|\Delta \tilde{T} |$ with respect to $\eta$, given by:
\begin{align}
    \frac{d \left| \Delta \tilde{T} \left( \eta \right) \right|}{d \eta} = \eta \frac{k^{2} v_{ud}^{2} \left( k^{2} v_{ud}^{2} + 2 \mu^{2} \right) - \left( \eta^{2} + \mu^{2} \right)^{2}}{\sqrt{\eta^{2}+\mu^{2}} \left( \left( k^{2} v_{ud}^{2} - \eta^{2} \right)^{2} + \left( \eta \mu \right)^{2} \right)^{3/2}}
\end{align}
also vanishes at a non-zero temporal frequency given by:
\begin{equation*}
    \eta_{\pm} = \pm \sqrt{ - \mu^{2} + \sqrt{ k^{2} v_{ud}^{2} \left( k^{2} v_{ud}^{2} + 2 \mu^{2} \right) }}  \label{eq:oc_roots}
\end{equation*}
which is real-valued when:
\begin{align*}
    k^{2} v_{ud}^{2} \left( k^{2} v_{ud}^{2} + 2 \mu^{2} \right) & > \mu^{4} \\
    \implies 1 + 8 \chi_{OC} - 16 \chi_{OC}^{2} & > 0 \\
    \implies \chi_{OC} < \frac{\sqrt{2} + 1}{4} & \approx 0.6.
\end{align*}
thus resulting in additional local extrema at $\eta_{\pm}$. Since in the right neighborhood of $\eta = 0$ ($\eta = \delta \eta$), $| \Delta \tilde{T} |$ has a positive slope when $\chi_{OC} < 0.6$, the local extremum at $\eta_{+}$ is, in fact, a global maximum of $|\Delta \tilde{T} |$, as shown in Fig.~\ref{fig:schematic_fourier_analysis}~(b). Thus, for $\chi_{OC} < 0.6$, the magnitude of the transient temperature response in the frequency domain exhibits a single peak with a non-zero width at a non-zero frequency, $\eta_+$, indicating the presence of coherent, albeit decaying, temporal fluctuations of the temperature field, thus resulting in the damped-oscillatory collective second sound regime of hydrodynamic phonon transport in real space, as described in the main text. Therefore, the oscillatory classifier, $\chi_{OC} < 1$ in Eq.\ref{eq:OC_v1} is updated to $\chi_{OC} < 0.6$ in Eq.~\ref{eq:visibility_condition}. On the other hand, for $\chi_{OC} > 0.6$, the extrema occur only at $\eta = 0$ and $\eta \to \infty$, thus resulting in a monotonically decreasing function, as shown in Fig.~\ref{fig:schematic_fourier_analysis} (b).

This frequency domain analysis is pivotal in distinguishing the decaying oscillatory response due to the hydrodynamic second sound from the ballistic heat flow. While the negative temperature dip in the time domain temperature response in the TG experimental geometry is indeed a signature of hydrodynamic second sound, consistent with the literature (e.g., see Refs.~\cite{huberman_observation_2019, ding_observation_2022}) and presented in Fig.~\ref{fig:bte_to_hyperbolic_equation} (schematic) and Fig.~\ref{fig:tg_soln_Graphene} (LPBE solution for graphene at 100~K for a TG period of 50~$\mu$m) of this manuscript, we have shown in Fig.~\ref{fig:beyond_ddc_tg_soln_Graphene} of the main text for graphene that such time domain features can also occur in the ballistic heat flow regime. However, the distinction between the two regimes is clear in the frequency-domain solution ($| \Delta \tilde{T} |$), with the hydrodynamic second sound exhibiting one broad peak at non-zero $\eta$, as shown in Fig.~\ref{fig:freq_domain_tg_soln_Graphene} for graphene at 100~K and a TG period of 50~$\mu$m, whereas several narrow peaks are observed in the inset of Fig.~\ref{fig:beyond_ddc_tg_soln_Graphene} for graphene at 100~K and a TG period of 0.1~$\mu$m, representing ballistic motion of phonons with different group velocities.

\section{Limiting hydrodynamic temperature response at long heating length scales} \label{app_sec:hha_at_large_grating}
In the main text, we emphasized that at long heating length scales, represented by large grating periods ($k=2 \pi / d \rightarrow 0$) in a TG set up, the predicted solution of the HHE is indistinguishable from that of the FHE, indicating that the observed temperature response is Fourier-diffusive in nature. To demonstrate this feature, we begin with the approximation:
\begin{gather*}
        \sqrt{D'} = \sqrt{\mu^{2} - 4 v_{ud}^{2} k^{2}} \approx \mu \left( 1 - \frac{2 v_{ud}^{2} k^{2}}{\mu^{2}} \right) \\
        \frac{\mu}{\sqrt{D'}} \rightarrow 1
\end{gather*}
at large $d$. From the solution of HHE for $D>0$ [Eq.~\ref{eq:non_oscilatory_second_sound_TG_time_domain_temperature_response}], the spatio-temporal temperature response in the frequency domain becomes:
\begin{align}
    \Delta T \left( x, t \right) & = \Delta T_{0} e^{-ikx} e^{- \mu t / 2} u \left( t \right) \nonumber \\
    \times & \left[ \frac{\mu}{\sqrt{D'}} \sinh{\left( \frac{\sqrt{D'}}{2}t \right)} + \cosh{\left( \frac{\sqrt{D'}}{2}t \right)} \right] \\
    & \approx \Delta T_{0} e^{-ikx} e^{- \mu t / 2} u \left( t \right) \nonumber \\
    \times & \left[ \sinh{\left( \frac{\sqrt{D'}}{2}t \right)} + \cosh{\left( \frac{\sqrt{D'}}{2}t \right)} \right] \\
    & = \Delta T_{0} e^{-ikx} e^{- \mu t / 2} u \left( t \right) \exp \left( \frac{\sqrt{D'}}{2}t \right)\nonumber \\
    & \approx \Delta T_{0} e^{-ikx} e^{- \mu t / 2} u \left( t \right) \exp \left( \frac{\mu t}{2} - \frac{v_{ud}^{2} k^{2}}{\mu}t \right)\nonumber \\
    & = \Delta T_{0} e^{-ikx} e^{- v_{ud}^{2} k^{2} t / \mu} u \left( t \right)
\end{align}
which resembles the solution of FHE with the decay rate $v_{ud}^{2} k^{2} / \mu$.

\begin{acknowledgments}
This work was supported by the Core Research Grant (CRG) No. CRG/2022/009160, and the Mathematical Research Impact Centric Support (MATRICS) grant no. MTR/2022/001043 from the Department of Science and Technology - Science and Engineering Research Board, India, by the Advanced Research Grant (ARG) No. ANRF/ARG/2025/007160/ENS from the Anusandhan National Research Foundation, India and by the Infosys Foundation through a Young Investigator Award (N.K.R.). N.M. gratefully acknowledges the Prime Minister's Research Fellowship (PMRF) grant no. PMRF-02-01036. The authors acknowledge Pragyesh Sangal for useful discussions.
\end{acknowledgments}

\section*{Author contributions}
N.K.R. originated the research idea. N.M. developed the computational framework and performed the calculations. N.M. and N.K.R. developed the theory, analyzed the results, and wrote the manuscript.

\section*{Code availability}
All formulations and computational optimizations necessary to perform the calculations presented in this manuscript are described in the Methods section and in refs.~\cite{ravichandran_unified_2018, malviya_failure_2023}.


\bibliography{references}

\end{document}